# Triplet J-driven DNP – a proposal to increase the sensitivity of solution-state NMR without microwave

Maria Grazia Concilio[1*], Yiwen Wang[1], and Linjun Wang[2,3], Xueqian Kong[1,2*]

[1]*Institute of Translational Medicine, Shanghai Jiao Tong University, 200240 Shanghai, China*

[2]*Department of Chemistry, Zhejiang University, 310058 Hangzhou, Zhejiang, China*

[3]*Zhejiang Key Laboratory of Excited-State Energy Conversion and Energy Storage, Department of Chemistry, Zhejiang University, 310058 Hangzhou, Zhejiang, China*

**Abstract**

Dynamic nuclear polarization (DNP) is an important method to enhance the limited sensitivity of nuclear magnetic resonance (NMR). Using the existing mechanisms such as Overhauser DNP (ODNP) is still difficult to achieve significant enhancement of NMR signals in solutions at a high magnetic field. The recently proposed J-driven DNP (JDNP) condition (when the exchange interaction $J_{ex}$ of two electron spins matches the electron or the nuclear Lamour frequency $\omega_E$ and $\omega_N$) may enable signal enhancement in solution as it requires only dipolar interaction between the biradical polarization agent and the analyte. However, likewise ODNP, the current JDNP strategy still requires the saturation of the electron polarization with high microwave power which has poor penetration and is associated with heating effects in most liquids. The replacement of high-power microwave irradiation is possible if the temporal electron polarization imbalance is created by an electron electromagnetic (EM) irradiation at different wavelength such as the visible light. Here, we propose a triplet-JDNP mechanism which first exploits the light-induced singlet fission process (i.e., a singlet exciton is converted into two triplet excitons). As the JDNP condition $J_{ex} \approx \pm\omega_E$ is fulfilled, a triplet-to-triplet cross-relaxation process will occur with different rates and consequently lead to the creation of hyperpolarization on the coupled nuclear spin states. This communication discusses the theory behind the triplet-JDNP proposal, as well as the polarizing agents and conditions that will enable the new approach to enhance NMR's sensitivity without the need of microwave irradiation.

*mariagrazia.concilio@sjtu.edu.cn

*xkong@sjtu.edu.cn

## Introduction:

Nuclear magnetic resonance (NMR) is one of the most versatile forms of spectroscopy, conveying structure and dynamics information with minimal invasiveness. Further applications of NMR are limited by its poor sensitivity, due to the low thermal polarization of nuclear spins. Methods based on dynamic nuclear polarization (DNP), [1-4] which transfer polarization from an unpaired electron in a radical polarizing agent to nearby nuclei, can alleviate the sensitivity problem. DNP has achieved remarkable success in solid-state NMR which greatly enhances the signals of low abundance nuclei and the sparse signals on the interfaces [5, 6]. For solution-state NMR, DNP relies on the Overhauser mechanism (mainly based on the Fermi contact coupling between unpaired electron and the target nuclear) [7-14] to achieve polarization transfer from unpaired electrons to their coupled nuclei such as $^{31}P$,[15-17] $^{19}F$ [7, 18] ,$^{13}C$. [8, 9, 11, 14, 19, 20] and in some cases also $^{1}H$ [8, 21]. So far, Overhauser DNP (ODNP) has been used to study metabolites [9, 11] amino acids [13, 21] and lipids [22, 23] in various solutions. [20] However, existing DNP techniques for both solid-state NMR and solution-state NMR require the use of high-power microwave irradiation to saturate the electron spins. The reliance of microwave irradiation is problematic due to the sample heating effect and the low microwave penetration, in particular for liquid samples. In addition, the efficiency of ODNP decreases significantly with the magnetic field limiting its applications in nowadays' high-field solution-state NMR. On the other hand, the microwave sources for DNP-relevant frequencies (from tens to hundreds of GHz) are yet satisfactory as they are costly, deficient in power, unable to generate pulses. [15]

We have theoretically proposed the J-driven DNP (JDNP) mechanism which works for solution-state NMR at various magnetic fields. The JDNP approach relies on the use of a coupled electron spin pair (on a biradical) whose exchange coupling $J_{ex}$ matches either the electron or the nuclear Larmor frequency, $\omega_E$ and $\omega_N$, respectively. The JDNP effect is different from ODNP, because it arises from intermolecular dipolar hyperfine coupling (between $^{1}H$ nuclei and biradical), while it does not require the presence of Fermi contact coupling (i.e. the scalar hyperfine coupling). [24, 25] The physics of the JDNP is reminiscent of that observed in chemically induced dynamic nuclear polarization (CIDNP) [26-29] in which nuclear magnetization enhancement is generated in a spin-correlated radical pair due to a singlet/triplet interconversion process modulated by dipolar hyperfine couplings. In the JDNP, microwaves (rather than the photochemical reactions in CIDNP) are used to drive the system away from the thermal equilibrium. As one of the JDNP conditions (either $J_{ex} \approx \pm\omega_E$ or $J_{ex} \approx \pm\omega_N$) is fulfilled, triplet-to-singlet cross-relaxation rates will occur with different rates for α and β nuclear spin states, and consequently lead to the creation of longitudinal nuclear polarization. [25] Meanwhile, a different approach named triplet-DNP is currently under development which also achieved notable signal enhancement in both solids and liquids. [30-32] [33-37] It is based on the photoexcitation of triplet transitions on a fluorescent molecule that creates temporal electron spin polarization. Nevertheless, the current triplet-DNP still needs a microwave source to facilitate the polarization transfer, referred as the integrated solid effect (ISE). [30, 38]

Bearing in mind that the reliance of microwave irradiation is the Achilles' heel of today's DNP techniques, we propose a fundamentally different DNP strategy which could potentially eliminate the use of microwave. The proposed strategy combines the underlying principles of both JDNP and triplet-DNP (dubbed triplet-JDNP) and it can generate nuclear spin polarization in the liquid state with only photoexcitation. The triplet-JDNP approach is based upon the singlet fission (SF) process of a triplet-pair system (e.g. on a chromophore dimer) in which a singlet exciton is converted into two triplet excitons by photoexcitation. [39-41] When $J_{ex}$ of the two triplets fulfills the JDNP condition $J_{ex} \approx \pm\omega_E$, the population imbalance may be generated on the α and β nuclear spin states (on the analyte molecule) that are coupled to the triplet pair through dipolar hyperfine interaction.

In a typical four-stage intramolecular SF model (Figure 1), for linked dimers composed by two pentacene chromophores, [42, 43] the absorption of a photon brings one chromophore from its ground state ($\hat{S}_0$) to its lowest-lying excited singlet configuration ($\hat{S}_1$). The resulting photogenerated singlet exciton $\hat{S}_1\hat{S}_0$ decays into correlated triplet pair states of multiplicity M=2S+1, where S is the total spin of the triplet pair state. The state $\hat{S}_1\hat{S}_0$ (or $\hat{S}_0\hat{S}_1$) corresponds to one pentacene in its first excited state, while the other remains in its ground state. In the case of a symmetric molecule, like the pentacene dimer considered in this study, the state $\hat{S}_1\hat{S}_0$ is equivalent to the state $\hat{S}_0\hat{S}_1$. These states can be labelled as ${}^{M}(\hat{T}\hat{T})_{m_s}$, where $m_s$ is the projection of spin angular moment. In the last stage of the SF, the triplet pair dissociates into two isolated triplets $\hat{T}_1$ + $\hat{T}_1$ which later decay away through a recombination process.

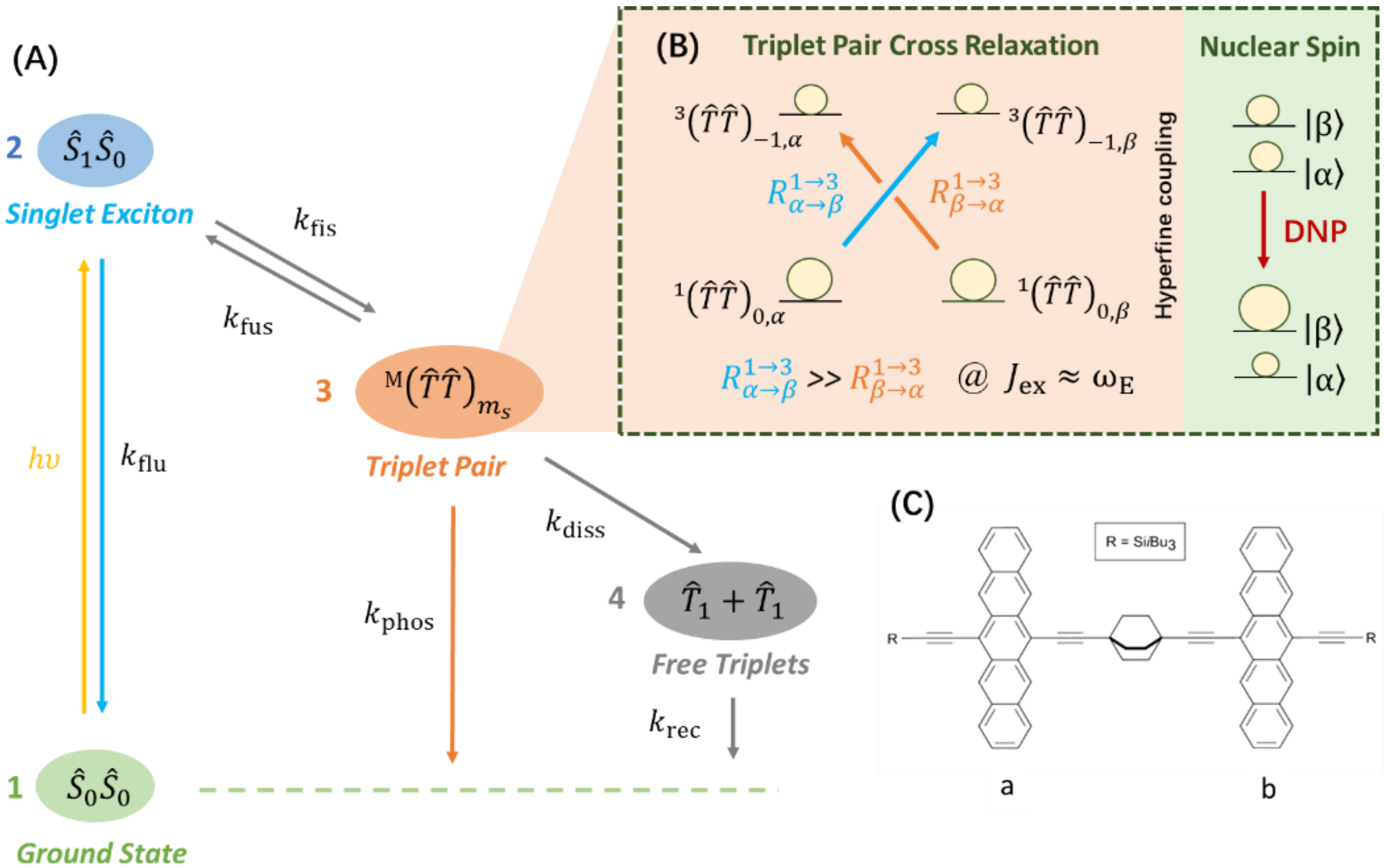

**Figure 1.** (A) The singlet fission scheme, where the constants $k_{\text{flu}}$, $k_{\text{fis}}$, $k_{\text{fus}}$, $k_{\text{diss}}$, $k_{\text{phos}}$ and $k_{\text{rec}}$ represent the rates of fluorescence, fission, fusion, dissociation, phosphorescence and recombination respectively. The numbers indicate the 4 stages of the SF process. (B) At the JDNP condition (when $J_{\text{ex}} \approx \omega_E$ is fulfilled), the triplet-to-triplet cross-relaxation process takes place with different rates from $^1(\hat{T}\hat{T})_{0,\alpha}$ to $^3(\hat{T}\hat{T})_{-1,\beta}$ (labeled as $R^{1\to3}_{\alpha\to\beta}$) and from $^1(\hat{T}\hat{T})_{0,\beta}$ to $^3(\hat{T}\hat{T})_{-1,\alpha}$ (labeled as $R^{1\to3}_{\beta\to\alpha}$). It leads to an increase of the population of the states $^3(\hat{T}\hat{T})_{-1,\alpha/\beta}$ and creates nuclear spin polarization of coupled protons. (C) The chemical structure of the pentacene dimer considered in this work which is composed of two monomers a and b.

In the case of a coupled triplet pair, there are nine possible spin states that can be constructed and labelled as $|S, m_S\rangle$, where S is equal to 2, 1 and 0. Five states with S = 2 and M = 5 are termed as $^5(\hat{T}\hat{T})_{0,\pm1,\pm2}$, three states with S = 1 and M = 3 are termed as $^3(\hat{T}\hat{T})_{0,\pm1}$, and one state with S = 0 and M = 1 is termed as $^1(\hat{T}\hat{T})_0$. The initial triplet pair formed by SF has overall singlet character and corresponds to the state $^1(\hat{T}\hat{T})_0$. Subsequently, the states $^3(\hat{T}\hat{T})_{0,\pm1}$ and $^5(\hat{T}\hat{T})_{0,\pm1,\pm2}$ are populated through relaxation or intersystem crossing (ISC). [42] For the analysis of JDNP condition, triplet pair states are coupled to nuclear spins (represented by a proton with α and β spin states) through dipolar hyperfine interaction. As $J_{\text{ex}} \approx \pm\omega_E$ is fulfilled at the JDNP condition, a population imbalance between the α and β nuclear components of the triplet pair states takes place (represented as $^1(\hat{T}\hat{T})_{0,\alpha/\beta}$, $^3(\hat{T}\hat{T})_{0,\pm1,\alpha/\beta}$ and $^5(\hat{T}\hat{T})_{0,\pm1,\pm2,\alpha/\beta}$) and hence hyperpolarization is created on the coupled protons. Nuclear polarization arises in particular from the states $^1(\hat{T}\hat{T})_{0,\alpha/\beta}$ and

$^3(\hat{T}\hat{T})_{\pm1,\alpha/\beta}$ (specifically $^3(\hat{T}\hat{T})_{-1,\alpha/\beta}$ in the case of positive $J_{ex}$ and $^3(\hat{T}\hat{T})_{+1,\alpha/\beta}$ in the case of negative $J_{ex}$). In this communication, we describe the mechanisms of triplet-JDNP using the Redfield theory [44] and also discuss the potential polarizing agents and conditions for enhancement optimization.

**Spin system and simulation methodology:**

A non-conjugated pentacene dimer with structure shown in Figure 1 was considered as the model system in this work. [42, 43] The singlet fission process involves states corresponding to the singlet ground state, the singlet exciton, the triplet pair states and the isolated triplet states. The dynamics of these excited states is usually modelled with either the Lindblad or the Bloch-Redfield master equation. [45] However, these formalisms do not support relaxation arising from tensor anisotropies from which the triplet-JDNP arises. Therefore, we used the Liouville space formulation of Bloch–Redfield–Wangsness (BRW) relaxation theory. [46, 47] Within the BRW formalism, it is not possible to create basis sets that describe the electronic states simultaneously in the same spin space. In addition, the triplet-JDNP effect arises only from the triplet pair states coupled to nuclear spin states. Therefore, our time domain simulations were performed using an approach similar to the one used to simulate the magnetic field effects in photoluminescence and CIDNP. [48, 49] We built a spin system composed by a triplet pair with two spin-1 and one proton spin-1/2. The proton and the triplet pair interact solely through anisotropic dipolar hyperfine couplings (scalar hyperfine couplings were not included); while the electrons in the triplet pair interact through inter-electron exchange coupling and triplet-triplet dipolar interactions. As in JDNP, we observed a constant and spherical symmetric distribution of the enhancement in the region surrounding the triplet-dimer. [24] A proton can get polarized in a distance from 0.5 Å up to 5 Å away from each monomer. Therefore, in our simulations, distances between the two triplets (belonging to the radical) and the proton (belonging to an assumed solvent molecule) were set to 3 Å and 17 Å respectively (the actual proton and triplet Cartesian coordinates are reported in Table 1), that corresponds to a proton in the region 0.5 Å - 5 Å away from each monomer. In order to simulate the SF process, we used a system of four coupled Liouville von Neumann equations to describe the time evolution of different states (see Eq. (3) later). In the numerical implementation, the Liouvillian is a supermatrix and the initial condition corresponds to a state vector containing the states $\hat{S}_0\hat{S}_0$, $\hat{S}_1\hat{S}_0$, $^{\mathrm{M}}(\hat{T}\hat{T})_{\mathrm{m_s}}$ and $\hat{T}_1 + \hat{T}_1$. The electron $g$-tensors were set with the eigenvalues taken from those found in pentacene in a single crystal of naphthalene, [50, 51] and the zero field splitting parameters D and E were set equal to 1139 MHz and 60 MHz as those observed in pentacene dimers. [52, 53] The rotational correlation time of the biradical/proton dipolar hyperfine coupled triad was set to 300 ps, as it should be longer than the rotational correlation time determined for the 6,13-bis(triisopropylsilylethynyl)-pentacene (TIPS)-pentacene. [54] Numerical simulations took into account of all self- and cross-relaxation terms within the BRW relaxation theory, [44, 47] and considered the local vibrational relaxation modes arising from the mixing between spin and orbital angular momentum (which often dominates the electron relaxation rates of organic radicals). [54] Spectral density functions describing the relaxation rates were computed with the pattern matching algorithm. [47] The kinetics constants $k_{\mathrm{flu}}$, $k_{\mathrm{fis}}$, $k_{\mathrm{fus}}$, $k_{\mathrm{diss}}$, $k_{\mathrm{phos}}$ and $k_{\mathrm{rec}}$ involved in the SF model shown in Figure 1 were determined experimentally using transient absorption spectroscopy and were taken from ref. [42]. All the simulation parameters are summarized in Table 1. Time domain simulations were performed with the Spinach software. [55]

**Table 1:** Parameters for the triplet-pair / proton spin system used in the simulations.

| **Parameter** | **Value** |
|---|---|
| $^1$H chemical shift tensor, ppm | [5 5 5] |
| Triplet 1 $g$-tensor eigenvalues, [xx yy zz] / Bohr magneton* | [2.0015 2.0009 2.0005] |
| Triplet 1 $g$-tensor, ZYZ active Euler angles / rad | [0.0 0.0 0.0] |
| Triplet 2 $g$-tensor eigenvalues, [xx yy zz] / Bohr magneton* | [2.0015 2.0009 2.0005] |
| Triplet 2 $g$-tensor, ZYZ active Euler angles / rad | [0.0 0.0 0.0] |
| Zero field splitting, D / MHz and E / MHz | 1139 and 60 |
| $^1$H coordinates [x y z] / Å | [0.0 0.0 10] |
| Triplet 1 and triplet 2 coordinates, [x y z] / Å | [0 0 -7] and [0 0 7] |
| Rotational correlation time $\tau_{\mathrm{C}}$ / ps | 300 |
| Kinetics constants, $k_{\mathrm{flu}}$, $k_{\mathrm{fis}}$, $k_{\mathrm{fus}}$, $k_{\mathrm{phos}}$, $k_{\mathrm{diss}}$ and $k_{\mathrm{rec}}$ / s$^{-1}$ | $9.5\times10^7$, $2.4\times10^9$, $5\times10^7$, $1\times10^7$, $5.7\times10^6$, $3\times10^4$ |
| Local mode relaxation / Hz | $1\times10^4$ |
| Temperature / K | 298 |

*Simulations used Zeeman interaction tensors that were made axially symmetric along the main molecular axis (corresponding for instance, to a linker connecting two pentacene units).

The nine triplet pair states can be represented in terms of the two spin-1 bases [56]:

$$\begin{gathered} {}^5(\hat{T}\hat{T})_{\pm2} = |\hat{T}_\pm\hat{T}_\pm\rangle,\ {}^5(\hat{T}\hat{T})_{\pm1} = \frac{|\hat{T}_0\hat{T}_\pm\rangle + |\hat{T}_\pm\hat{T}_0\rangle}{\sqrt{2}},\ {}^5(\hat{T}\hat{T})_0 = \frac{|\hat{T}_-\hat{T}_+\rangle + 2|\hat{T}_0\hat{T}_0\rangle + |\hat{T}_+\hat{T}_-\rangle}{\sqrt{6}}, \\ {}^3(\hat{T}\hat{T})_{\pm1} = \frac{|\hat{T}_0\hat{T}_\pm\rangle - |\hat{T}_\pm\hat{T}_0\rangle}{\sqrt{2}},\ {}^3(\hat{T}\hat{T})_0 = \frac{|\hat{T}_+\hat{T}_-\rangle - |\hat{T}_-\hat{T}_+\rangle}{\sqrt{2}} \\ {}^1(\hat{T}\hat{T})_0 = \frac{|\hat{T}_-\hat{T}_+\rangle - |\hat{T}_0\hat{T}_0\rangle + |\hat{T}_+\hat{T}_-\rangle}{\sqrt{3}} \end{gathered} \tag{1}$$

The triplet pair states combined with the proton spin states α and β correspond to:

$$^{5}\left(\hat{T}\hat{T}\right)_{\pm 2,\alpha/\beta} = \left|\hat{T}_{\pm}\hat{T}_{\pm}\alpha/\beta\right\rangle,\ ^{5}\left(\hat{T}\hat{T}\right)_{\pm 1,\alpha/\beta} = \frac{\left|\hat{T}_{0}\hat{T}_{\pm}\alpha/\beta\right\rangle + \left|\hat{T}_{\pm}\hat{T}_{0}\alpha/\beta\right\rangle}{\sqrt{2}},\ ^{5}\left(\hat{T}\hat{T}\right)_{0,\alpha/\beta} = \frac{|\hat{T}_{-}\hat{T}_{+}\alpha/\beta\rangle + 2|\hat{T}_{0}\hat{T}_{0}\alpha/\beta\rangle + |\hat{T}_{+}\hat{T}_{-}\alpha/\beta\rangle}{\sqrt{6}}$$

$$^{3}\left(\hat{T}\hat{T}\right)_{\pm 1,\alpha/\beta} = \frac{|\hat{T}_{0}\hat{T}_{\pm}\alpha/\beta\rangle - |\hat{T}_{\pm}\hat{T}_{0}\alpha/\beta\rangle}{\sqrt{2}},\ ^{3}\left(\hat{T}\hat{T}\right)_{0,\alpha/\beta} = \frac{|\hat{T}_{+}\hat{T}_{-}\alpha/\beta\rangle - |\hat{T}_{-}\hat{T}_{+}\alpha/\beta\rangle}{\sqrt{2}} \tag{2}$$

$$^{1}\left(\hat{T}\hat{T}\right)_{0,\alpha/\beta} = = \frac{|\hat{T}_{-}\hat{T}_{+}\alpha/\beta\rangle - |\hat{T}_{0}\hat{T}_{0}\alpha/\beta\rangle + |\hat{T}_{+}\hat{T}_{-}\alpha/\beta\rangle}{\sqrt{3}}$$

Triplet pair states are constructed with the Clebsch-Gordon coefficients in the Hilbert space [57] and irreducible spherical tensor operator (ISTO) are used in the time domain Liouville space simulations. In the later context, the state symbols are also used to indicate the corresponding state amplitude. ISTO of the states given in Eqs. 1 and 2 were determined with the software SpinDynamica (https://www.spindynamica.soton.ac.uk/) and were then implemented into the software Spinach [55]. The SF process shown in Figure 1 can be represented with a set of coupled Liouville von Neumann equations:

$$\frac{\partial \hat{\rho}_1(t)}{\partial t} = k_{\text{flu}}\,\hat{\rho}_2(t) + k_{\text{phos}}\,\hat{\rho}_3(t) + k_{\text{rec}}\,\hat{\rho}_4(t)$$

$$\frac{\partial \hat{\rho}_2(t)}{\partial t} = -(k_{\text{flu}} + k_{\text{fis}})\,\hat{\rho}_2(t) + k_{\text{fus}}\ ^{1}\left(\hat{T}\hat{T}\right)_{0,L}\hat{\rho}_3(t)\ ^{1}\left(\hat{T}\hat{T}\right)_{0,R}$$

$$\frac{\partial \hat{\rho}_3(t)}{\partial t} = k_{\text{fis}}\,\hat{\rho}_2(t) - i\left[\hat{\hat{L}}, \hat{\rho}_3(t)\right] - \frac{k_{\text{fus}}}{2}\left[^{1}\left(\hat{T}\hat{T}\right)_{0}, \hat{\rho}_3(t)\right]_{+} - \left(k_{\text{diss}} + k_{\text{phos}}\right)\hat{\rho}_3(t) \tag{3}$$

$$\frac{\partial \hat{\rho}_4(t)}{\partial t} = k_{\text{diss}}\,\hat{\rho}_3(t) - k_{\text{rec}}\,\hat{\rho}_4(t)$$

where $\hat{\rho}_{1-4}(t)$ correspond to the states $\hat{S}_0\hat{S}_0$, $\hat{S}_1\hat{S}_0$, $^{M}\left(\hat{T}\hat{T}\right)_{ms} = {}^{M}\left(\hat{T}\hat{T}\right)_{ms/\alpha} + {}^{M}\left(\hat{T}\hat{T}\right)_{ms/\beta}$ and $\hat{T}_1 + \hat{T}_1$ respectively; $^{1}\left(\hat{T}\hat{T}\right)_{0,L}$ and $^{1}\left(\hat{T}\hat{T}\right)_{0,R}$ are the left- and right-side product superoperators of the state $^{1}\left(\hat{T}\hat{T}\right)_{0}$; $\left[^{1}\left(\hat{T}\hat{T}\right)_{0}, \hat{\rho}_3(t)\right]_{+} = {}^{1}\left(\hat{T}\hat{T}\right)_{0,L}\hat{\rho}_3(t) + \hat{\rho}_3(t)\,{}^{1}\left(\hat{T}\hat{T}\right)_{0,R}$ is the anticommutation superoperator; $\hat{\hat{L}}$ is the Liouvillian superoperator. Eq. (3) can be numerically implemented by building a joint Liouville space of the four sites, all the operators were built considering a spin system composed by two spin-1 and one proton spin-1/2. The Liouvillian superoperator becomes a supermatrix equal to:

$$\hat{\hat{L}} = \hat{\hat{H}} + \hat{\hat{R}} + \hat{\hat{K}} \tag{4}$$

where $\hat{\hat{H}}$, $\hat{\hat{R}}$ and $\hat{\hat{K}}$ are the spin Hamiltonian supermatrix, the relaxation supermatrix and the kinetics supermatrix, respectively. They correspond to:

$$\hat{\hat{H}} = \begin{pmatrix} \hat{0} & \hat{0} & \hat{0} & \hat{0} \\ \hat{0} & \hat{0} & \hat{0} & \hat{0} \\ \hat{0} & \hat{0} & \hat{\hat{H}}_{\mathrm{TT}} & \hat{0} \\ \hat{0} & \hat{0} & \hat{0} & \hat{0} \end{pmatrix}, \hat{\hat{R}} = \begin{pmatrix} \hat{0} & \hat{0} & \hat{0} & \hat{0} \\ \hat{0} & \hat{0} & \hat{0} & \hat{0} \\ \hat{0} & \hat{0} & \hat{\hat{R}}_{\mathrm{TT}} & \hat{0} \\ \hat{0} & \hat{0} & \hat{0} & \hat{0} \end{pmatrix},$$

$$\hat{\hat{K}} = \begin{pmatrix} \hat{0} & k_{\mathrm{flu}}\hat{1} & k_{\mathrm{phos}}\hat{1} & k_{\mathrm{rec}}\hat{1} \\ \hat{0} & -(k_{\mathrm{flu}} + k_{\mathrm{fis}})\hat{1} & k_{\mathrm{fus}}\left[ {}^{1}(\hat{T}\hat{T})_{L} \cdot {}^{1}(\hat{T}\hat{T})_{R} \right] & \hat{0} \\ \hat{0} & k_{\mathrm{fis}}\hat{1} & -\frac{k_{\mathrm{fus}}}{2}\left[ {}^{1}(\hat{T}\hat{T})_{L} + {}^{1}(\hat{T}\hat{T})_{R} \right] - (k_{\mathrm{diss}} + k_{\mathrm{phos}})\hat{1} & \hat{0} \\ \hat{0} & \hat{0} & k_{\mathrm{diss}}\hat{1} & -k_{\mathrm{rec}}\hat{1} \end{pmatrix} \tag{5}$$

where $\hat{\hat{H}}_{\mathrm{TT}}$ is the laboratory frame spin Hamiltonian that contains all the interactions present in the coupled spin system of the triplet pair and the proton, while $\hat{\hat{R}}_{\mathrm{TT}}$ is the relaxation superoperator that contains all the relaxation mechanisms arising from tensor anisotropies and relaxation from local vibrational modes. $\hat{0} = \begin{pmatrix} 0 & \cdots & 0 \\ \vdots & \ddots & \vdots \\ 0 & \cdots & 0 \end{pmatrix}$ and $\hat{1} = \begin{pmatrix} 1 & \cdots & 0 \\ \vdots & \ddots & \vdots \\ 0 & \cdots & 1 \end{pmatrix}$ are the zero and the unit matrices respectively with the same dimension of $\hat{\hat{R}}_{\mathrm{TT}}$ and $\hat{\hat{H}}_{\mathrm{TT}}$. The laboratory frame Hamiltonian corresponds to:

$$\begin{aligned} \hat{\hat{H}} &= \hat{\hat{H}}_{\mathrm{Z,a}} + \hat{\hat{H}}_{\mathrm{Z,b}} + \hat{\hat{H}}_{\mathrm{Z,N}} + \hat{\hat{H}}_{\mathrm{ZFS,a}} + \hat{\hat{H}}_{\mathrm{ZFS,b}} + \hat{\hat{H}}_{\mathrm{DD}} + \hat{\hat{H}}_{\mathrm{J}} + \hat{\hat{H}}_{\mathrm{HF}} \\ &= \sum_{\mathrm{k=a,b}} \hat{\vec{S}}^{(k)} \cdot \mathbf{G}_{\mathbf{E}}^{(\mathbf{k})} \cdot \hat{\vec{\mathrm{B}}}_0 + \hat{\vec{\mathrm{N}}} \cdot \mathbf{G}_{\mathbf{N}} \cdot \hat{\vec{\mathrm{B}}}_0 + \sum_{\mathrm{k=a,b}} \hat{\vec{S}}^{(k)} \cdot \mathbf{Z}^{(\mathbf{k})} \cdot \hat{\vec{S}}^{(k)} + \\ &\quad \hat{\vec{S}}^{(a)} \cdot (\mathbf{D} + J_{\mathrm{ex}}) \cdot \hat{\vec{S}}^{(b)} + \sum_{\mathrm{k=a,b}} \hat{\vec{S}}^{(k)} \cdot \mathbf{A}^{(\mathbf{k})} \cdot \hat{\vec{\mathrm{N}}} \end{aligned} \tag{6}$$

where $\hat{\hat{H}}_{\mathrm{Z,a}}$, $\hat{\hat{H}}_{\mathrm{Z,b}}$ and $\hat{\hat{H}}_{\mathrm{Z,N}}$ represent the Zeemann interaction of the triplets in the monomers a and b and of the nuclear spin respectively, $\hat{\hat{H}}_{\mathrm{ZFS,a}}$ and $\hat{\hat{H}}_{\mathrm{ZFS,b}}$ represent the ZFS interaction of the triplet in the two monomers a and b respectively, $\hat{\hat{H}}_{\mathrm{DD}}$ represents the inter-electron dipolar interaction, $\hat{\hat{H}}_{\mathrm{J}}$ represents the inter-electron exchange coupling and $\hat{\hat{H}}_{\mathrm{HF}}$ represents the dipolar hyperfine interaction. $\hat{\vec{S}}^{(k)}$ with $k$ = a and b is a vector with the electron spin operators for the triplets in monomers a and b, $\hat{\vec{\mathrm{N}}}$ is a vector with the nuclear spin operators, $\hat{\vec{\mathrm{B}}}_0$ is the magnetic field vector, $\mathbf{G}_{E}^{(k)}$ and $\mathbf{Z}^{(\mathrm{k})}$ are the electron Zeeman and ZFS interaction tensors, $\mathbf{G}_{N}$ is the nuclear Zeeman interaction tensor, $\mathbf{D}$ is the inter-electron dipolar interaction tensor and $\boldsymbol{A}^{(\boldsymbol{k})}$ is the dipolar hyperfine interaction tensor between the nucleus and the indicated electron $k$. The Redfield relaxation superoperator and the analytical expressions of the cross- and self-relaxation rates involving two states a and b were computed in according to:

$$\left\langle \hat{\rho}_a \middle| \hat{\hat{R}}_{ab} \middle| \hat{\rho}_b \right\rangle = \sum_{kmpq} \int_0^{\infty} G_{kmpq}(\tau) \mathrm{Tr}\left( \hat{\rho}_a^{\dagger} \left[ \hat{Q}_{km}, \left[ e^{i\hat{H}_0\tau} \hat{Q}_{pq}^{\dagger} e^{-i\hat{H}_0\tau}, \hat{\rho}_b \right] \right] \right) d\tau \tag{7}$$

where $\hat{\hat{R}}_{ab}$ is the relaxation superoperator, $G_{kmpq}(\tau)$ is the correlation function between two Wigner functions that perform the ensemble average [47]; $\hat{H}_0$ is the static part of the Hamiltonian; $\hat{Q}_{km(pq)}$ are the rotational basis operators of the anisotropic part of the Hamiltonian. [44, 47] Time domain simulations were performed using the Liouville von Neumann equation of motion:

$$\frac{\partial\hat{\rho}(t)}{\partial t} = \hat{\rho}(0)\exp\left(-\hat{\hat{L}}t\right) \tag{8}$$

where $\hat{\rho}(0) = [\hat{\rho}_1(0), \hat{\rho}_2(0), \hat{\rho}_3(0), \hat{\rho}_4(0)]$ and $\hat{\rho}(t) = [\hat{\rho}_1(t), \hat{\rho}_2(t), \hat{\rho}_3(t), \hat{\rho}_4(t)]$ are supervectors containing the state vectors of four species involved in the SF at the time zero and at the time $t$ respectively. Simulations were performed by setting the initial condition to $\hat{\rho}(0)$=[0,1,0,0], corresponding to the $\hat{S}_1\hat{S}_0$, state being fully populated, while the populations of the other states were set to zero. The $\hat{S}_1\hat{S}_0$ state was set equal to ${}^1\big(\hat{T}\hat{T}\big)_0$ representing the triplet states with singlet character, in the simulations of the SF process, it is moved in the triplet pair subspace of the joint Liouville space in according to the $k_{\mathrm{fis}}$ and $k_{\mathrm{fus}}$ rates due to the ${}^1\big(\hat{T}\hat{T}\big)_{0,L}$ and ${}^1\big(\hat{T}\hat{T}\big)_{0,R}$ operators in Eq. 3. Time evolution of the observable operator $\hat{O}$, corresponding to the population of triplet pair states, was calculated by taking their scalar products with the state vector $\hat{\rho}_n(t)$, with n=1-4 (corresponding to the vectors in Eq. (3)), at each time step:

$$\hat{O}(t) = \langle \hat{O} | \hat{\rho}_n(t) \rangle \tag{9}$$

The time dynamics of $\hat{S}_0\hat{S}_0(t)$, $\hat{S}_1\hat{S}_0(t)$ and $\hat{T}_1 + \hat{T}_1(t)$ was computed by summing the amplitude of all the triplet pair states ${}^{\mathrm{M}}\big(\hat{T}\hat{T}\big)_{ms,\alpha/\beta}$, computed in according Eq. (9), using the vectors $\hat{\rho}_1(t)$, $\hat{\rho}_2(t)$ and $\hat{\rho}_4(t)$ in Eq. (3). In our kinetics model, we did not take into account of the correlated singlet–triplet pair states, since they were not observed in the time-resolved electron paramagnetic resonance (TREPR) spectrum of the molecule under consideration. [42, 43] Singlet–triplet pair states are expected to arise from the states ${}^3\big(\hat{T}\hat{T}\big)_{0,\pm1}$, in molecules where heavy atoms are present. [58] The states ${}^3\big(\hat{T}\hat{T}\big)_{0,\pm1}$ are commonly generated in spin system that contain heavy atoms with intersystem crossing due to the spin-orbit coupling. In the pentacene dimers we did not observe the specie ${}^3\big(\hat{T}\hat{T}\big)_{0,\pm1}$ in time resolved TREPR spectra because the spin-orbit coupling is negligible in this organic molecule. On the other hand, the states ${}^3\big(\hat{T}\hat{T}\big)_{0,\pm1}$ can be also created due to the self-relaxation rates within the Redfield theory, but their creation happens on timescales longer than the overall singlet fission process. Simulations were performed from 0 T to 14.1 T. The evolution of $k_{\mathrm{fis}}$ and $k_{\mathrm{fus}}$ with the magnetic field can be predicted with the Merrifield model. [57, 59, 60] According to the Merrifield model, the kinetics constants $k_{\mathrm{fis}} = k_{\mathrm{fis}}\left|C^l_{\mathrm{singlet}}\right|^2$ and $k_{\mathrm{fus}} = k_{\mathrm{fus}}\left|C^l_{\mathrm{singlet}}\right|^2$ are dependent on the magnetic field due to the dependence of the singlet character $C^l_{\mathrm{singlet}} = \left\langle {}^1\big(\hat{T}\hat{T}\big)_0 \middle| \varphi_l \right\rangle$ where $\varphi_l$ with $l$ = 1-9 are the eigenstates of the spin Hamiltonian of the triplet pair at a given the magnetic field. [60] $C^l_{\text{singlet}}$ represents the amount of $^1(\hat{T}\hat{T})_0$ in the eigenstates. In solids, the anisotropic parts of the spin interactions lead to variations of $C^l_{\text{singlet}}$ with the magnetic field, but in liquids they are averaged out. Therefore, $C^l_{\text{singlet}}$ is set to 1 at the magnetic fields considered in this work, and consequently the SF kinetic rates are expected to remain constant with the magnetic field (see Figure S1 in the supporting material for more details).

## Results and discussion:

### The photo-kinetics of Triplet-JDNP:

The inter-electron exchange coupling $J_{ex}$ of the pentacene dimer shown in Figure 1 is about 15 GHz. This value was used to simulate the TREPR experiments on pentacene dimers. [53] Therefore, we performed triplet-JDNP simulations at 0.55 T to fulfill the JDNP condition $J_{ex} \approx \omega_E$ = 15 GHz. The simulation results are shown in Figure 2.

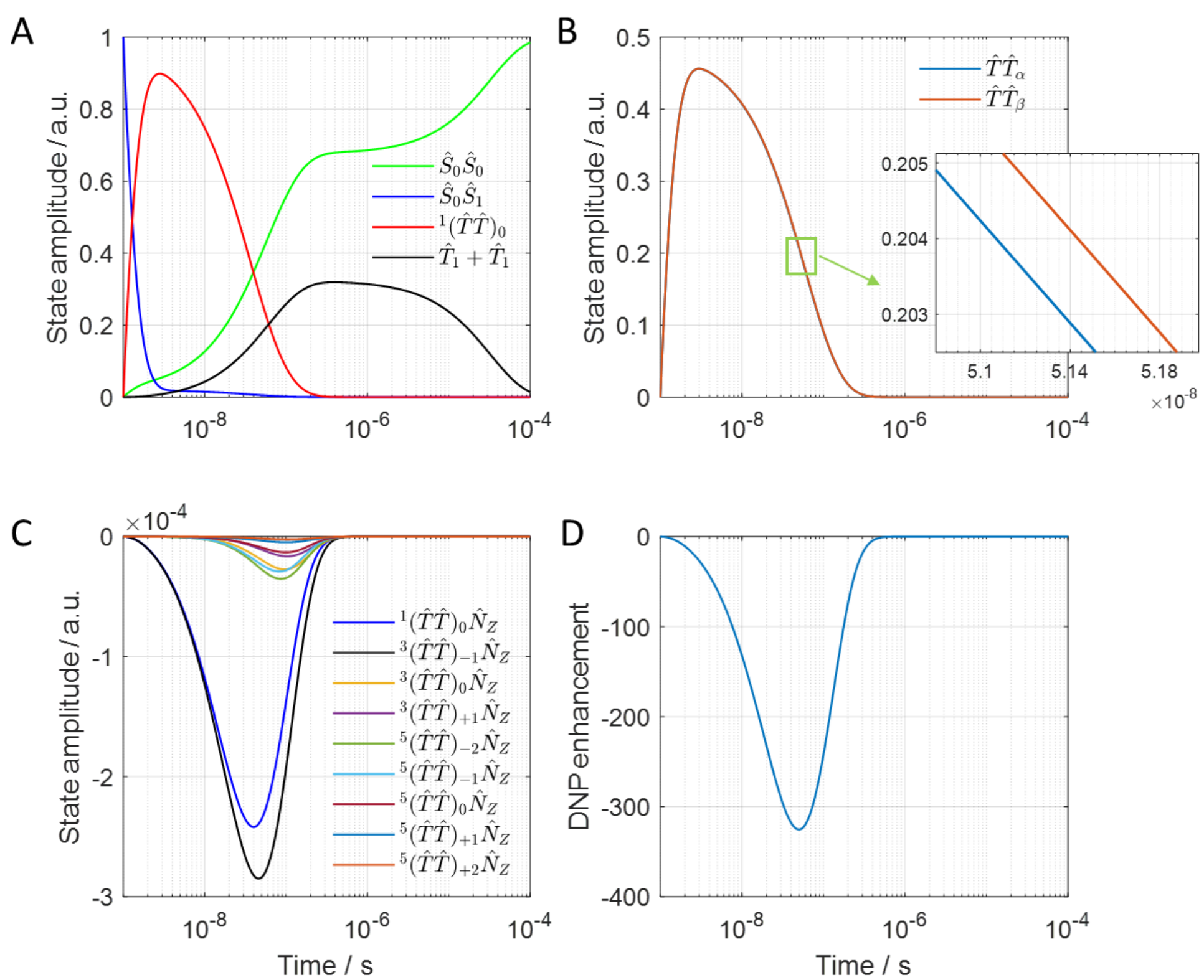

**Figure 2.** (A) Time domain simulations performed at 0.55 T, showing the evolution of the singlet ground state, singlet exciton, the triplet pair state with singlet character and the dissociated triplet states. (B) The sum of the α and β nuclear

components of the nine triplet pair states, indicated as $\hat{T}\hat{T}_{\alpha}$ and $\hat{T}\hat{T}_{\beta}$. (C) States arising from the imbalance between the α and β components of the triplet pair states, indicated with the states ${}^{\mathrm{M}}\big(\hat{T}\hat{T}\big)_{m_s}\hat{N}_Z$, where $\hat{N}_Z$ represents the nuclear magnetization. (D) $^{1}$H DNP enhancement at different time points. The DNP enhancement represents the achieved nuclear polarization, normalized by its Boltzmann counterpart at the same temperature and field. The other simulation parameters are given in Table 1.

The time evolution of the states shown in Figure 2A is consistent with that observed experimentally using transient optical absorption spectroscopy. [42, 43] At thermal equilibrium, the pentacene dimer exists in its singlet ground state, represented by the state $\hat{S}_0\hat{S}_0$. There is a negligible difference between the populations of the α and β nuclear states, Figure 2B. After the laser irradiation, the singlet exciton $\hat{S}_1\hat{S}_0$ is created (set as time zero in our simulations). Afterwards, the singlet exciton is converted to the state ${}^{1}\big(\hat{T}\hat{T}\big)_0$ with a fission rate $k_{\mathrm{fis}}$. The states ${}^{3}\big(\hat{T}\hat{T}\big)_{0,\pm1}$ and ${}^{5}\big(\hat{T}\hat{T}\big)_{0,\pm1,\pm2}$ are then populated due to the relaxation mechanisms described by the relaxation superoperator. Subsequently, the triplet pair states dissociate into free triplets $\hat{T}_1$+$\hat{T}_1$ (which do not contribute to nuclear polarization under the JDNP mechanism), and eventually decay to the ground states $\hat{S}_0\hat{S}_0$. The lifetimes of $\hat{S}_1\hat{S}_0$, the triplet pair states and the dissociated triplets are about 0.004 μs, 0.4 μs and 99 μs respectively. The triplet pair states are coupled to the nuclear spin states α and β (indicated as ${}^{\mathrm{M}}\big(\hat{T}\hat{T}\big)_{m_s,\alpha/\beta}$ ). The time evolution of the sum of α and β nuclear components of the nine triplet pair states, represented as $\hat{T}\hat{T}_{\alpha}$ and $\hat{T}\hat{T}_{\beta}$, is shown in Figure 2B. As the JDNP condition is fulfilled, [24, 25] a population imbalance between ${}^{\mathrm{M}}\big(\hat{T}\hat{T}\big)_{m_s/\alpha}$ and ${}^{\mathrm{M}}\big(\hat{T}\hat{T}\big)_{m_s/\beta}$ states takes place, populating the states ${}^{\mathrm{M}}\big(\hat{T}\hat{T}\big)_{m_s}\hat{N}_Z$ (Figure 2C). The nuclear polarization is created from the population difference between the states $\hat{T}\hat{T}_{\alpha}$ and $\hat{T}\hat{T}_{\beta}$ in according to the relation:

$$\hat{N}_Z = \hat{T}\hat{T}_{\alpha} - \hat{T}\hat{T}_{\beta} \tag{10}$$

where:

$$\hat{T}\hat{T}_{\alpha/\beta} = \frac{{}^{1}(\hat{T}\hat{T})_{0,\alpha/\beta} + {}^{3}(\hat{T}\hat{T})_{0,\alpha/\beta} + {}^{3}(\hat{T}\hat{T})_{\pm1,\alpha/\beta} + {}^{5}(\hat{T}\hat{T})_{0,\alpha/\beta} + {}^{5}(\hat{T}\hat{T})_{\pm1,\alpha/\beta} + {}^{5}(\hat{T}\hat{T})_{\pm1,\alpha/\beta} + {}^{5}(\hat{T}\hat{T})_{\pm2,\alpha/\beta}}{2} \tag{11}$$

The build-up of the nuclear polarization is a transient event that lasts about 0.5 μs and reaches its maximum 0.05 μs after the laser irradiation (see Figure 2D). In the case of positive $J_{\mathrm{ex}}$, the nuclear polarization arises mostly from an imbalance between the α and β nuclear components of the ${}^{1}\big(\hat{T}\hat{T}\big)_0$ and ${}^{3}\big(\hat{T}\hat{T}\big)_{-1}$states. In the case of negative $J_{\mathrm{ex}}$, the state ${}^{3}\big(\hat{T}\hat{T}\big)_{+1}$ is involved instead of the state ${}^{3}\big(\hat{T}\hat{T}\big)_{-1}$. The "polarization region" in which the triplet JDNP is active is when the electron-nuclear distances are within 5 Å. In an actual experiment, the solvent proton diffuses randomly at $\approx$10 Å/μs (*i.e.* 0.001 μm/μs with a diffusion constant of $10^{-12}$ m$^2$ ms$^{-1}$ at 25 °C), [61] crossing several times in-and-out of the "polarization region". Considering a solvent proton located at 3 Å away from one of the triplets, it will take about 0.2 μs to become sufficiently polarized and then move out of the "polarization region" into the bulk where the nuclear relaxation is longer. As such, the polarization of the proton is enhanced and preserved.

**The polarization mechanism behind the triplet-JDNP:**

As the JDNP condition is fulfilled, the energies of the states, including $^1(\hat{T}\hat{T})_0$ and $^3(\hat{T}\hat{T})_{-1}$, $^3(\hat{T}\hat{T})_0$ and $^5(\hat{T}\hat{T})_{-2}$, as well as $^3(\hat{T}\hat{T})_{+1}$ and $^5(\hat{T}\hat{T})_{-1}$, cross with each other respectively at 0.55 T (Figure 3A) . We observed a difference in the triplet-to-triplet cross-relaxation rates between the states $^1(\hat{T}\hat{T})_{0,\alpha} \rightarrow {}^3(\hat{T}\hat{T})_{-1,\beta}$ and $^1(\hat{T}\hat{T})_{0,\beta} \rightarrow {}^3(\hat{T}\hat{T})_{-1,\alpha}$ (Figure 3B) according to the Redfield theory [47].

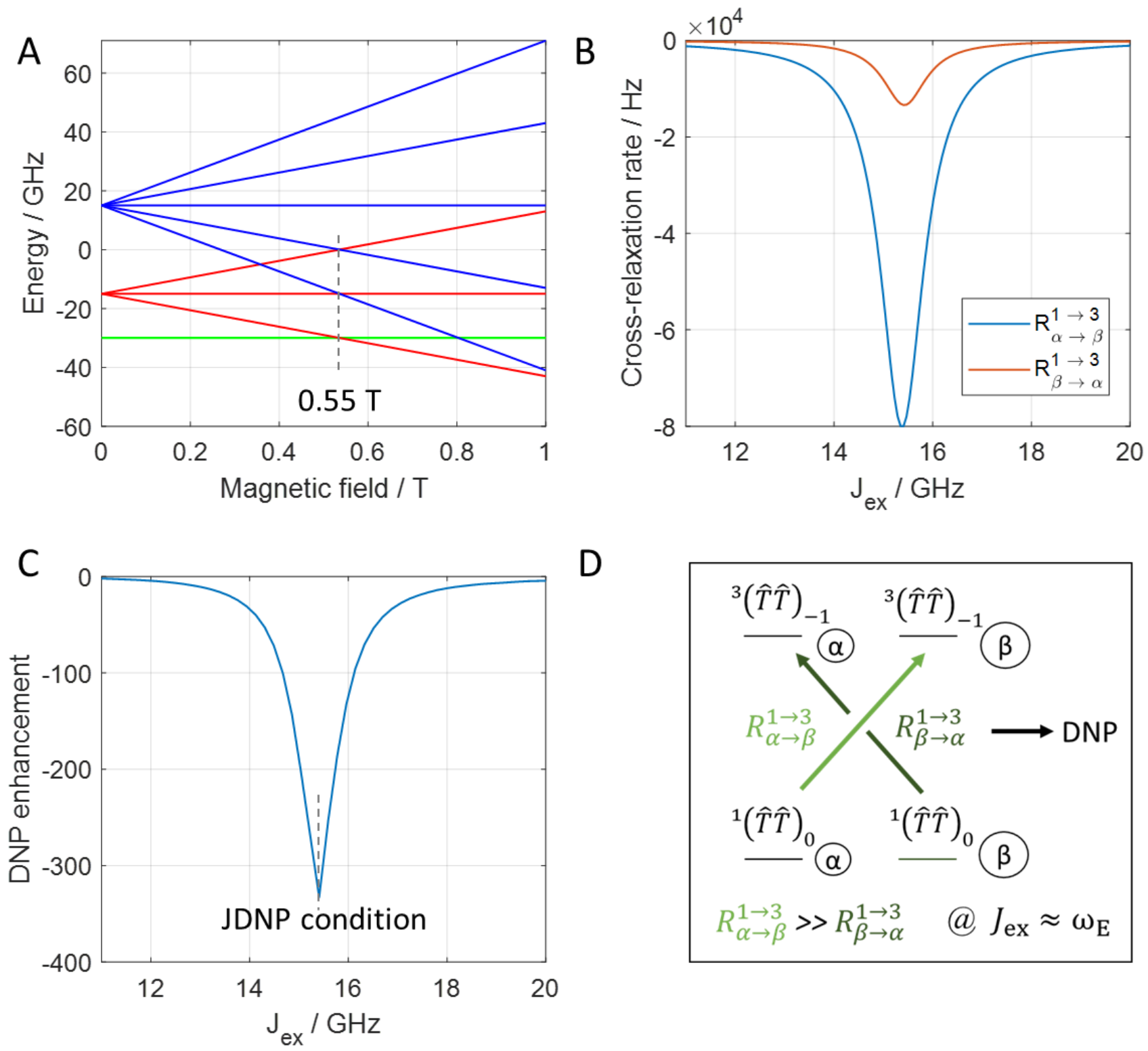

**Figure 3.** (A) Energy level diagram of the triplet pair states at different magnetic fields. The green line corresponds to the state $^1(\hat{T}\hat{T})_0$, the red lines correspond to the three states $^3(\hat{T}\hat{T})_{0,\pm1}$ and the blue lines correspond to the five states $^5(\hat{T}\hat{T})_{0,\pm1,\pm2}$. The dashed line indicates the level-crossing of energy levels at 0.55 T. (B) Triplet-to-triplet cross-relaxation rates $R^{1\rightarrow3}_{\alpha\rightarrow\beta}$ and $R^{1\rightarrow3}_{\alpha\rightarrow\beta}$, as functions of $J_{ex}$. (C) $^1$H DNP enhancement obtained as a function of $J_{ex}$. (D)

Diagram showing the states that cross-relax to each other with rates $R_{\alpha\to\beta}^{1\to3}$ and $R_{\beta\to\alpha}^{1\to3}$, leading to nuclear polarization $\hat{N}_Z$. The simulation parameters are given in Table 1.

The cross-relaxation rates between states ${}^1(\hat{T}\hat{T})_{0,\alpha/\beta}$ and ${}^3(\hat{T}\hat{T})_{-1,\beta/\alpha}$ can be described by the spectral density functions:

$$R_{\alpha\to\beta}^{1\to3} = \frac{4}{45}\Delta^2_{\mathrm{HFC_1-HFC_2}} J(J_{\mathrm{ex}} + \omega_{\mathrm{E}} + \omega_{\mathrm{N}}) \quad \text{for } {}^1(\hat{T}\hat{T})_{0,\alpha} \to {}^3(\hat{T}\hat{T})_{-1,\beta} \qquad (12)$$

and:

$$R_{\beta\to\alpha}^{1\to3} = \frac{2}{135}\Delta^2_{\mathrm{HFC_1-HFC_2}} J(J_{\mathrm{ex}} + \omega_{\mathrm{E}} - \omega_{\mathrm{N}}) \quad \text{for } {}^1(\hat{T}\hat{T})_{0,\beta} \to {}^3(\hat{T}\hat{T})_{-1,\alpha} \qquad (13)$$

where the term $\Delta^2_{\mathrm{HFC_1-HFC_2}}$ is the a second rank norm squared [62] arising from the difference of the anisotropy of the two dipolar hyperfine coupling tensor matrices, indicated with $HFC_1$ and $HFC_2$. $J(\omega)$ is the spectral density function at a frequency ω corresponding to $\tau_{\mathrm{C}}/(1+\tau_{\mathrm{C}}^2\omega^2)$ arising from the Fourier transform of the correlation function in Eq. 7, [44, 47] where $\tau_{\mathrm{C}}$ is the rotational correlation time. The difference between the two dipolar hyperfine tensor anisotropies creates a cross-relaxation path between the states ${}^1(\hat{T}\hat{T})_{0,\alpha/\beta}$ and ${}^3(\hat{T}\hat{T})_{-1,\beta/\alpha}$. Tensor anisotropies do not create cross-relaxation paths between ${}^3(\hat{T}\hat{T})_{0,\alpha/\beta}$ and ${}^5(\hat{T}\hat{T})_{-2,\beta/\alpha}$, or between ${}^3(\hat{T}\hat{T})_{+1,\alpha/\beta}$ and ${}^5(\hat{T}\hat{T})_{-1,\beta/\alpha}$. The difference between the cross-relaxation rates $\Delta R_{\mathrm{cross}} = R_{\alpha\to\beta}^{1\to3} - R_{\beta\to\alpha}^{1\to3} \neq 0$ arises due to the presence of different denominators of the spectral density functions describing the rates. $\Delta R_{\mathrm{cross}}$ leads to an imbalance between the α and β nuclear spin states and consequently to the DNP enhancement (Figure 3C), in according to the polarization mechanism shown in Figure 3D. We also analyzed the self-relaxation rates $R[{}^{\mathrm{M}}\hat{T}\hat{T}_{m_s,\alpha}]$ and $R[{}^{\mathrm{M}}\hat{T}\hat{T}_{m_s,\beta}]$ which arise due to tensor anisotropies (see supporting material for more details). The results showed that the self-relaxation rates do not contribute directly to the DNP enhancement but indirectly to increase the populations of the triplet pair states from the state ${}^1(\hat{T}\hat{T})_0$.

## Analysis of triplet-JDNP dynamics with different parameters:

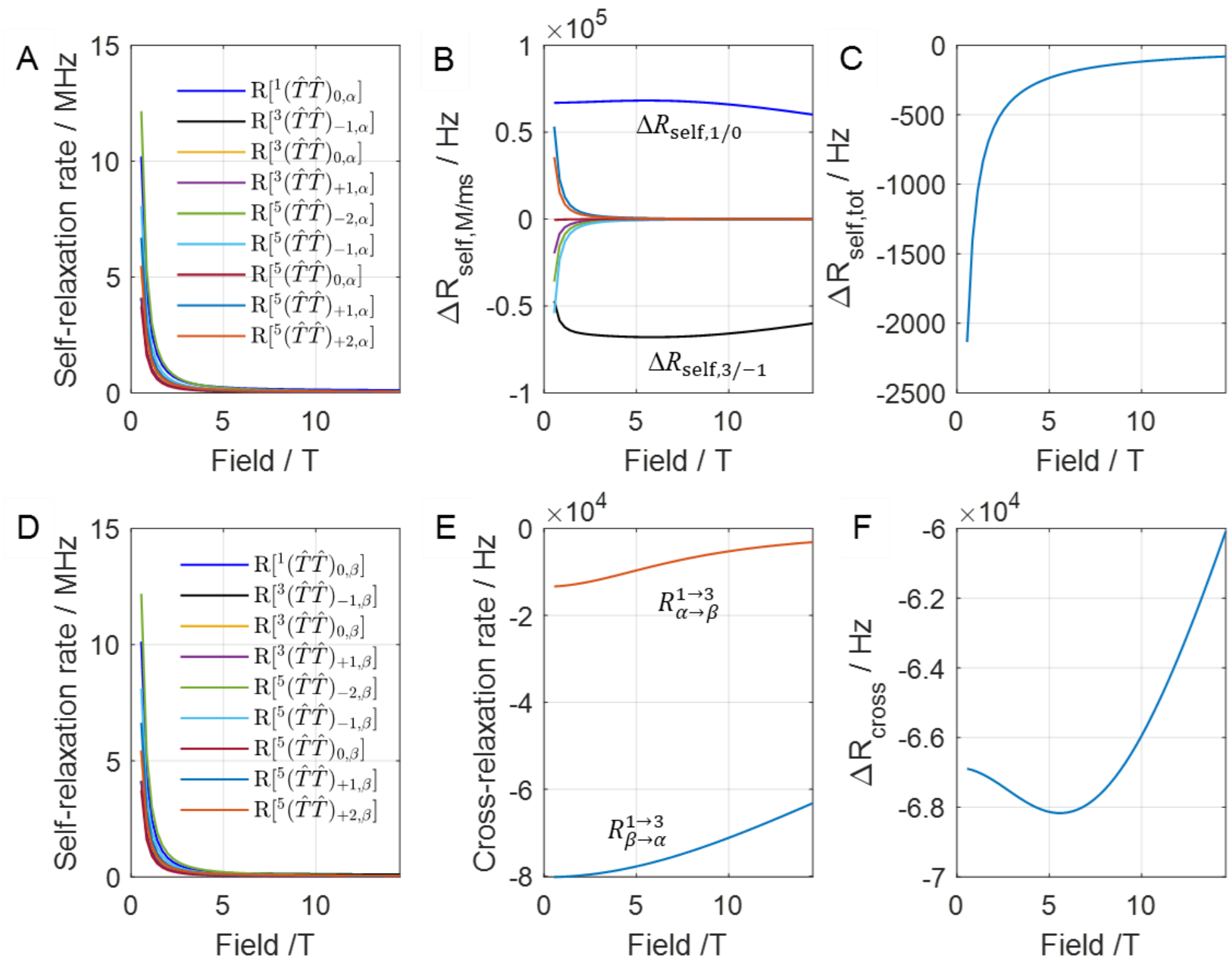

**Figure 4:** Self- and cross-relaxation rates as functions of the magnetic field. (A and D) Self-relaxation rates of the α and β nuclear components of the triplet pair states. (B) The difference of self-relaxation rates ($\Delta R_{self,M/ms}$) of the triplet pair states. The color code is the same used to indicate the triplet pair states in the figures (A) and (D). (C) The difference between the sum of self-relaxation rates of all the triplet pair states, indicated with $\Delta R_{self,tot}$. (E) The cross-relaxation rates $R^{1\to3}_{\alpha\to\beta}$ and $R^{1\to3}_{\beta\to\alpha}$ and (F) the difference between the cross-relaxation rates $\Delta R_{cross} = R^{1\to3}_{\alpha\to\beta} - R^{1\to3}_{\beta\to\alpha}$. $J_{ex}$ was set at the JDNP condition $J_{ex} \approx \omega_E$, and the other simulation parameters are given in Table 1.

We evaluated both the triplet-to-triplet cross- and self-relaxation rates as functions of the magnetic field (see Figure 4). Figures 4A and 4D show the decrease of the self-relaxation rates of the α and β nuclear components of the triplet pair states with the magnetic field. The $\Delta R_{self,M/ms}$ and $\Delta R_{self,tot}$, generally decay to zero with the magnetic field, except for $\Delta R_{self,1/0}$ and $\Delta R_{self,3/-1}$ associated with the states $^1(\hat{T}\hat{T})_0$ and $^3(\hat{T}\hat{T})_{-1}$ (see Figures 4B and 4C). Nevertheless, the self-relaxation processes contribute little to the nuclear polarization as the magnitude of total difference is two orders of magnitude smaller than that of the cross-relaxation. The $R^{1\to3}_{\alpha\to\beta}$ and $R^{1\to3}_{\beta\to\alpha}$ rates decrease slightly with the magnetic field (see Figures 4E and 4F) and the $\Delta R_{cross}$ shows a small negative maximum at ~ 5 T (for the parameter set used in the calculation).

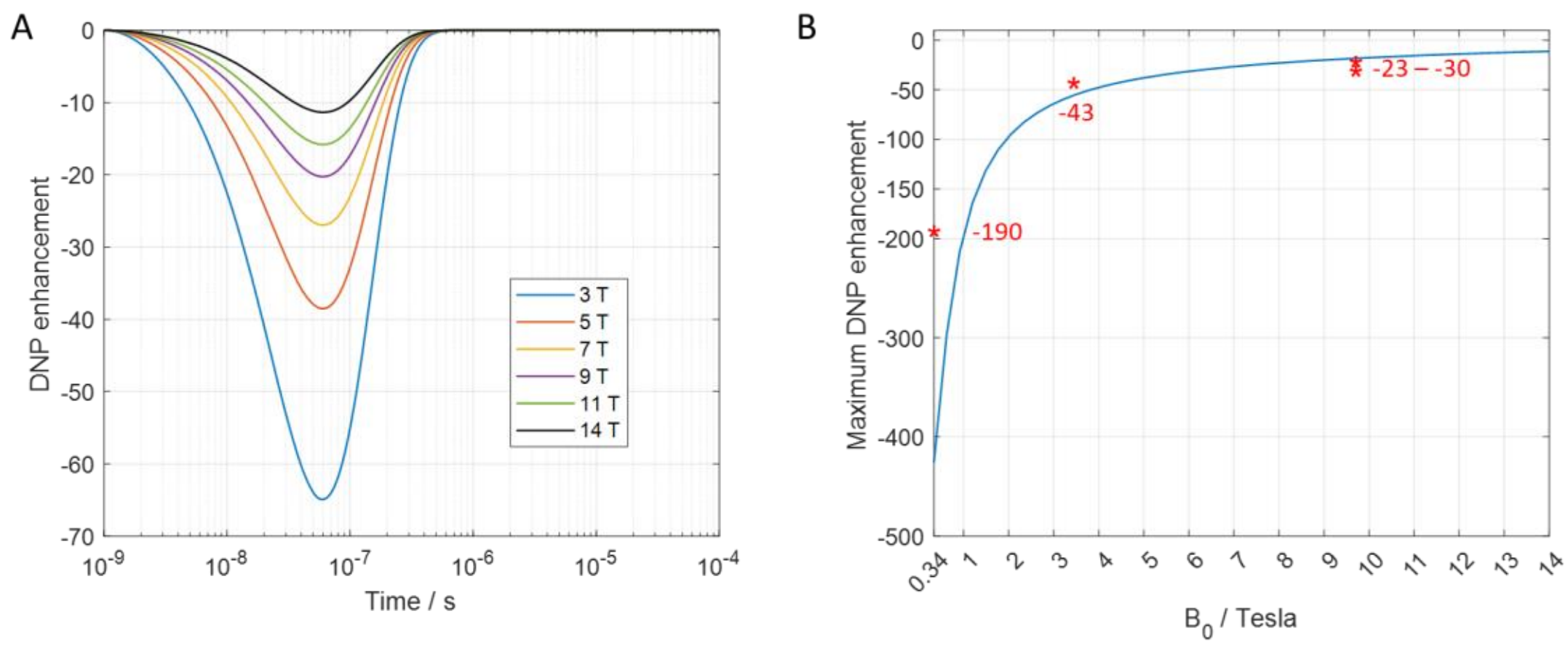

**Figure 5:** (A) Time-domain simulations showing the evolution of the [1]H DNP enhancement in time as a function of the magnetic field. (B) Comparison between the maximum enhancement obtained in the triplet-JDNP time-domain simulations (blue line) and that in steady-state [1]H ODNP experiments (red stars). $J_{ex}$ was set at the JDNP conditions $J_{ex} \approx \omega_E$, and the other simulation parameters are given in Table 1.

With the set photo-kinetics of the pentacene dimer, the DNP enhancement reaches its maximum at ~ 50 ns considering the kinetics rates given in Table 1 (Figure 5A). The trend of dynamics curves does not change with the magnetic field but their maximum reduces (Figure 5B). The enhancement factors of triplet-JDNP are comparable to those observed in [1]H ODNP performed at the same magnetic fields.

Once the state $^1(\hat{T}\hat{T})_0$ is created, the $\hat{T}\hat{T}_{\alpha/\beta}$ states are populated due to self-relaxation $R[^{\mathrm{M}}\hat{T}\hat{T}_{m_s,\alpha/\beta}]$. Subsequently, cross-relaxation rates $R^{1\rightarrow 3}_{\alpha\rightarrow\beta}$ and $R^{1\rightarrow 3}_{\beta\rightarrow\alpha}$ take place to create the population imbalance of the nuclear spin. Although the cross-relaxation rates remain constant with the magnetic field, the self-relaxation rates decay with the magnetic field. Thereby, the $\hat{T}\hat{T}_{\alpha/\beta}$ states will be less populated at the higher fields, and consequently the DNP enhancement is reduced.

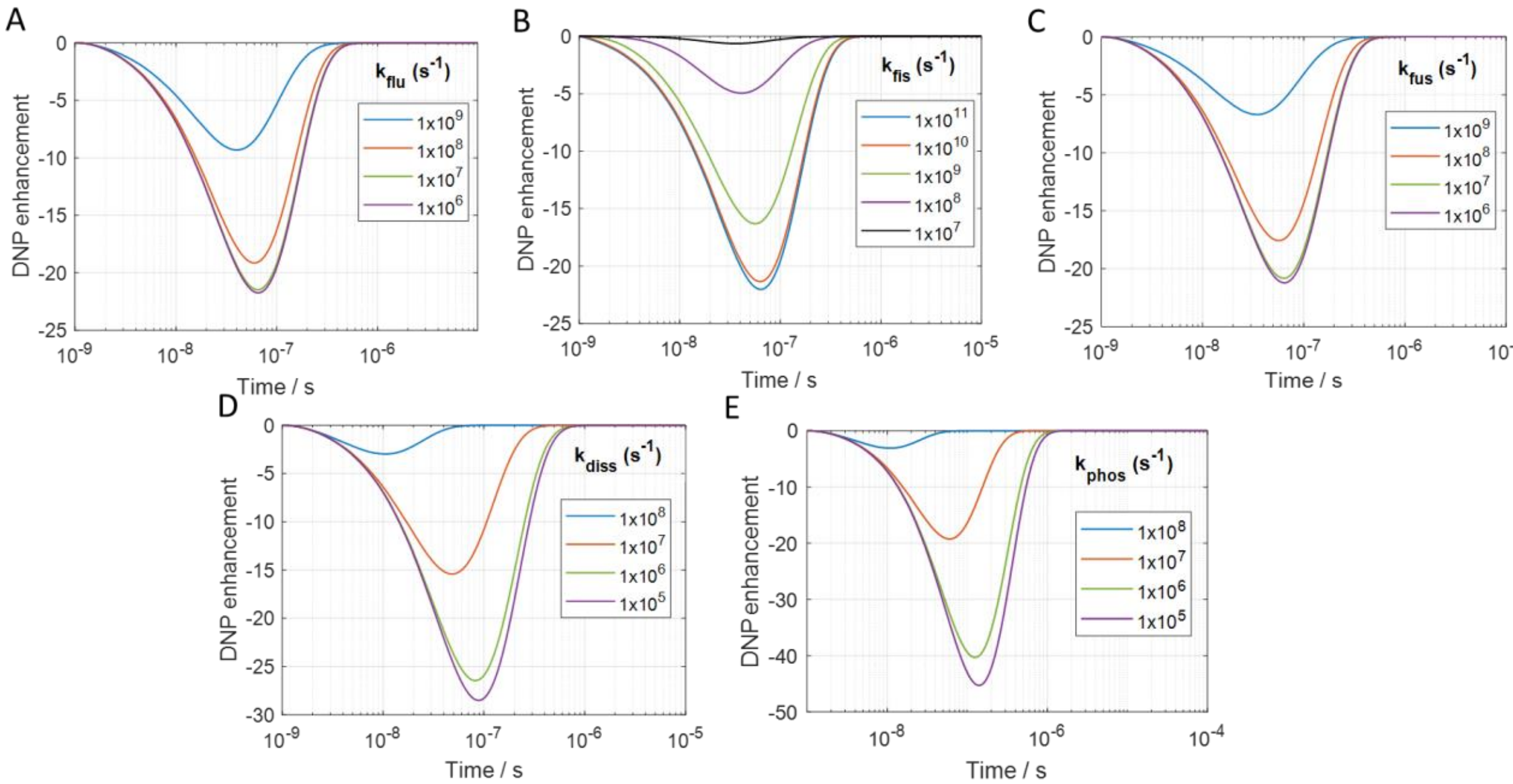

**Figure 6.** Effects of the kinetic rates on the DNP enhancement: (A) $k_{flu}$, (B) $k_{fis}$, (C) $k_{fus,}$ (D) $k_{diss}$ and (E) $k_{phos}$. Simulations were performed at 9.4 T. While one kinetics rate is changed, the other rates and parameters were kept the same as those shown in Table 1.

In order to provide insights for the design of an optimal polarizing agent for triplet-JDNP, we investigated the effects of the kinetics rates on the DNP enhancement. We analyzed the effect of the kinetics rates on a hypothetical pentacene dimer molecule that can fulfill the JDNP condition ($J_{ex} \approx \omega_E = 263$ GHz) at a high magnet field (e.g. 9.4 T). We also showed the effect of the kinetics rates as a function of the $J_{ex}$, see the supporting material in Figure S3. In Figure 6, we observed an increase of the DNP enhancement with the decrease of $k_{flu}$ that creates $\hat{S}_1\hat{S}_0$ exciton and an increase of $k_{fis}$ that generates the triplet pair state $^1(\hat{T}\hat{T})_0$. The DNP enhancement increases also with the decrease of $k_{diss}$, $k_{fus}$ and $k_{phos}$ which subtracts population from the triplet pair states. The favorable conditions can be found for $k_{flu} \leq 10^7$ s$^{-1}$, $k_{fis} \geq 10^9$ s$^{-1}$, $k_{diss} \leq 10^6$ s$^{-1}$, $k_{fus} \leq 10^7$ s$^{-1}$ and $k_{phos} \leq 10^6$ s$^{-1}$. In general, $k_{flu}$ of the potential molecule has to be of the same magnitude or lower than $k_{fis}$ to obtain a substantial population of $\hat{S}_1\hat{S}_0$ and $^1(\hat{T}\hat{T})_0$. In turn, $k_{fis}$, and the self-relaxation rates $\Delta R_{self,M/ms}$ have to be at least comparable to $k_{fus}$, $k_{diss}$ and $k_{phos}$ to increase the population of the triplet pair states contributing to the DNP enhancement. Under the considered range of photo-kinetic rates, and if the JDNP condition can be fulfilled at a selected magnetic field, triplet-JDNP may achieve DNP enhancements between 20~50 at 9.4 T and between 10~30 at 14 T (Figure 7).

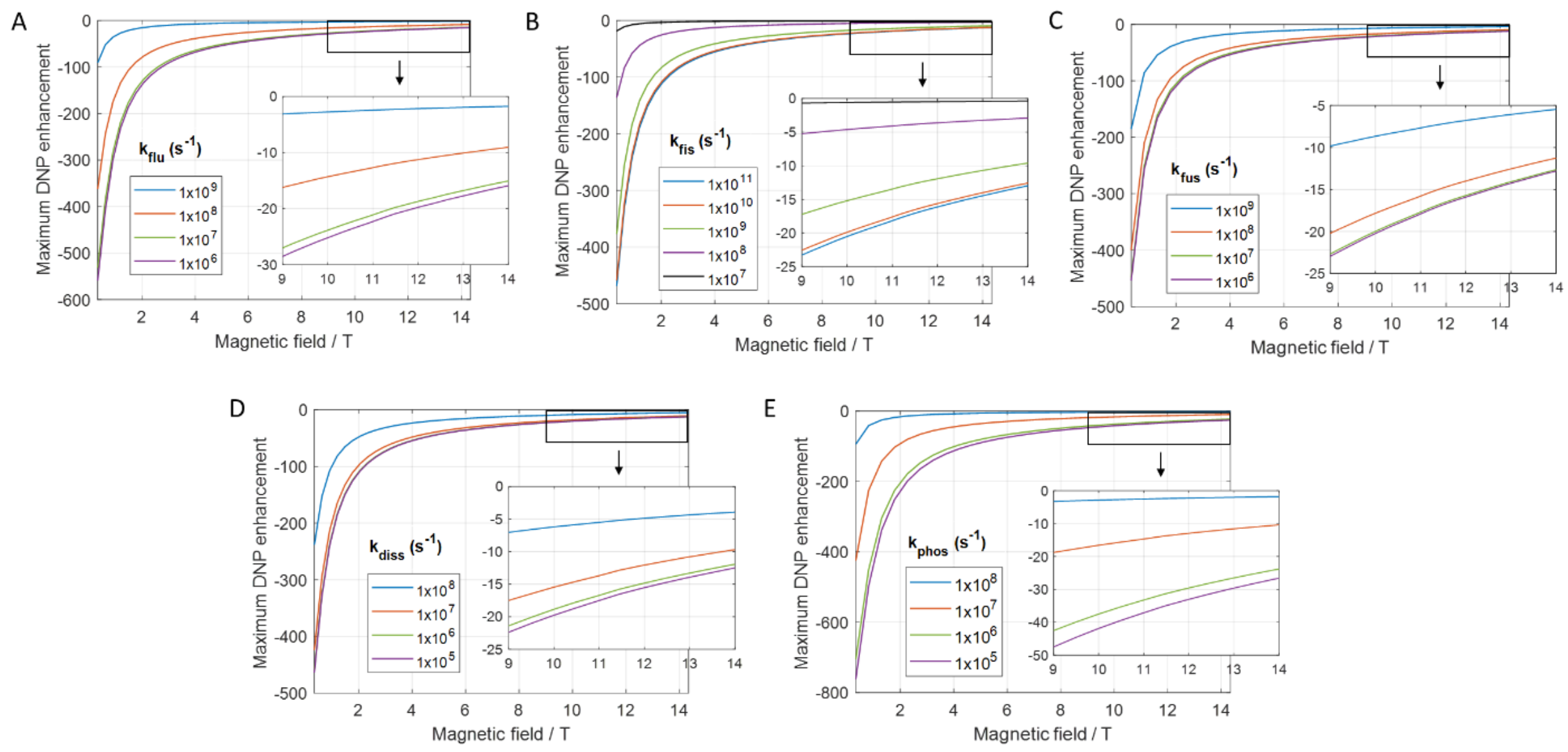

**Figure 7.** Maximum enhancement obtained in the triplet-JDNP simulations when $J_{ex}$ is set to match the JDNP condition under variable kinetic rates: (A) $k_{flu}$, (B) $k_{fis}$, (C) $k_{fus}$, (D) $k_{diss}$ and (E) $k_{phos}$. While one kinetics rate is changed, the other rates and parameters were kept the same as those shown in Table 1.

**Molecular systems for high-field triplet-JDNP:**

To realize triplet-JDNP at a high magnetic field (>3 T), the dimer molecule should have the following features:

1. An inter-electron exchange $J_{ex}$ in order of 100s GHz. Its strength depends on the overlap of the π orbitals of the two chromophores containing the triplet states. The presence of π orbitals in the linker (e.g conjugated linkers containing phenylene units) can mix efficiently with the π orbitals of the SF chromophores, leading to stronger $J_{ex}$(s) and in some cases also to prolonged triplet lifetimes. [53, 63] The strongest DNP enhancement can be observed at the condition $J_{ex} \approx \omega_E \pm 1$ GHz (see Figure S3 in the supporting material).

2. Chromophore dimers with better symmetry. Large differences in the ZFS- and g-tensor anisotropy leads to an increase of the electron relaxation that reduces the efficiency of polarization transfer. Axially symmetric or the nearly isotropic g-tensors, as well as the ZFS parameters, are expected to work for triplet-JDNP (an example is the TIPS-pentacene dimers). The ZFS anisotropy can be further reduced in biphenylene-based diradicals. [64] In dimers with identical chromophore monomers (Figure S6), the *g*-tensor non-collinearity arising from vibrational modes and rotations does not suppress the DNP enhancement.

3. Long-lived triplet pair states. In Figure 2, we observed that the nuclear polarization build-up reaches its maximum at about 50 ns. Therefore, triplet-JDNP requires long-lived triplet pair states whose lifetime is in the ns range (the longer is preferred). This can be achieved in flexible structures with high dynamic conformational space. [53] The presence an acetylene bond should induce mobility and significant structural heterogeneity in many linkers. [53]

4. Favorable SF rates. As shown in Figures 6 and 7, the favorable SF rates were fund to be $k_{flu} \leq 10^7$ $s^{-1}$, $k_{fis} \geq 10^9$ $s^{-1}$, $k_{diss} \leq 10^6$ $s^{-1}$, $k_{fus} \leq 10^7$ $s^{-1}$ and $k_{phos} \leq 10^6$ $s^{-1}$. The interplay between linker torsion and geometric fluctuations can be used to tune SF dynamics. [53] Kinetics rates, as well as the lifetime of the triplet pair states can be also manipulated by the solvent. [43, 53]

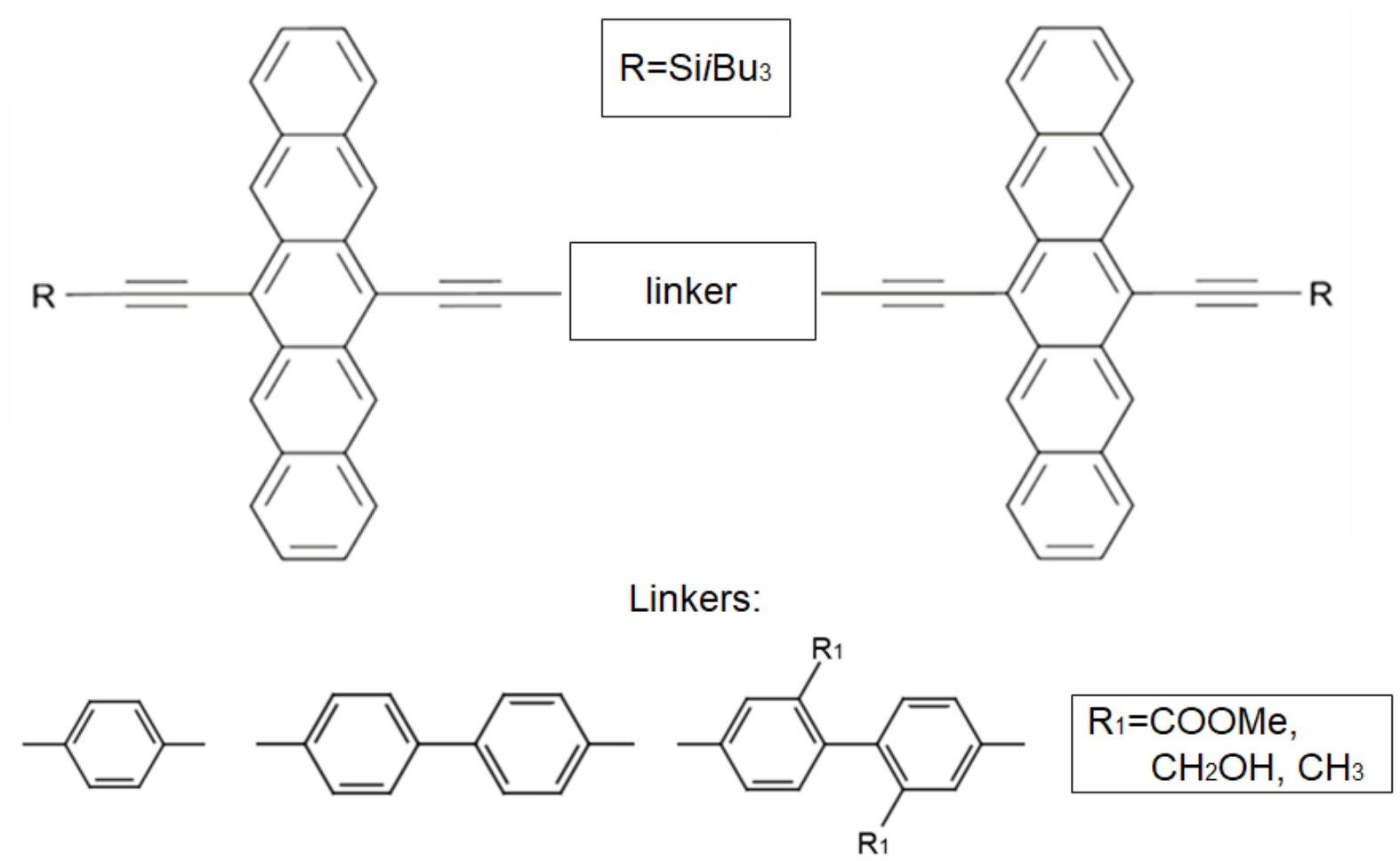

**Figure 8:** Chemical structures of TIPS-pentacene dimers bridged by phenylenes/acetylene linkers.

The above requirement can be archived in a variety of geometries and configurations of pentacene dimers bridged by phenylenes/acetylene linkers. Some have been already synthesized (see Figure 8). [53, 63] The inter-electron distance in the molecules shown in Figure 8 ranges between 12 Å -16 Å and the rotational correlation times are expected to range between about 150 - 300 ps. We did not observe a detectable variation of the DNP enhancement with the variation of the inter-electron distance between 12 Å -16 Å, while the DNP enhancement increases with the rotational correlation time (see Figure S7 in the supporting material).

**Conclusion and outlooks:**

We have explored the possibility of using pentacene dimers as polarizing agents for triplet-JDNP, which exploits the SF process to enhance the sensitivity of solution-state NMR's without microwaves. The simulated SF dynamics is consistent with that observed in transient absorption spectroscopy and TREPR experiments. At the JDNP condition ($J_{ex} \approx \omega_E$), there is a level anti-crossing between the states ${}^1(\hat{T}\hat{T})_0$ and ${}^3(\hat{T}\hat{T})_{-1}$ and triplet-to-triplet cross-relaxation takes place. A difference between these rates leads to an imbalance between the α and β nuclear components of the triplet pair states, and consequently to the creation of nuclear polarization. The DNP enhancement depends on the SF rates, and it can be enhanced with faster $k_{fis}$ that creates triplet pair states and slower $k_{flu}$, $k_{fus}$, $k_{diss}$, $k_{phos}$ that depopulate the triplet pair states. The SF process, although happening on the ns timescale, is sufficiently long to polarize a single solvent molecule.

In the experimental setup, the bulk solvent can be polarized by photoexciting the sample continuously using a laser of about 1 mW, which is similar to that used in triplet-DNP. [43, 65] Triplet-JDNP can be also performed at high magnetic fields (≥ 3 T), if $J_{ex}$ of the triplets can be increased. [66, 67] The expected advantage of the triplet-JDNP over the other hyperpolarization methods, including ODNP, CIDNP and parahydrogen-induced polarization (PHIP), [68] is the broader applicability to $^1H/^{13}C$ spectroscopy of a wider range of analytes, the possibility to use at higher magnetic fields, and the removal of microwave source. However, it still requires a JDNP condition to be fulfilled at a given field, and the optimal experimental conditions have to be determined. In the literature, there are many pentacene dimers that could suit for triplet-JDNP. [43, 53] The selection and optimization of various polarizing agents can be made trackable using transient optical absorption spectroscopy, photoluminescence experiments and TREPR. For instance, the inter-electron exchange coupling can be measured from the magnetic field effects obtained in steady-state fluorescence measurements performed on frozen solutions at a specific magnetic field. The SF process was used previously to enhance solid-state NMR's sensitivity. [69] However, the polarization mechanism is different from the one proposed for triplet-JDNP. It exploited the ISE [30, 38] under microwave irradiation to transfer polarization from the electrons to the nuclei, while triplet-JDNP is not expected to require microwave irradiation. In principle, the longer relaxation times of triplet pair and protons in solids should lead to a better DNP enhancement than that in liquids. We are currently investigating the possibility of applying triplet-JDNP principle to solid samples.

**Acknowledgements:**

This project is supported by the National Natural Science Foundation of China (Nos. 22425402, 22275159 and 22273082), the Fundamental Research Funds for the Central Universities (YG2025ZD30) and the Open Project Fund of National Facility for Translational Medicine (Shanghai) (TMSK-2024-109). The authors acknowledge Prof. Zhitao Zhang and Prof. Lu Yu (High Magnetic Field Laboratory of Chinese Academy of Sciences); Miss Linlin Ma and Miss Ruibin Wang (Shanghai Jiao Tong University); Prof. Haiming Zhu, and Dr. Shuangshuang Li (Zhejiang University) for the preliminary experiments and discussions.

**Supplementary material for**

# Triplet J-driven DNP – a proposal to increase the sensitivity of solution state NMR without microwave

Maria Grazia Concilio[1*], Yiwen Wang[1], and Linjun Wang[2,3], Xueqian Kong[1,2*]

[1]*Institute of Translational Medicine, Shanghai Jiao Tong University, 200240 Shanghai, China*

[2]*Department of Chemistry, Zhejiang University, 310058 Hangzhou, Zhejiang, China*

[3]*Zhejiang Key Laboratory of Excited-State Energy Conversion and Energy Storage, Department of Chemistry, Zhejiang University, 310058 Hangzhou, Zhejiang, China*

## Contents

### A. Magnetic field effect on the SF kinetics:

To determine the magnitude of the kinetics constants $k_{\mathrm{fis}} = k_{\mathrm{fis}} \left|C^{l}_{\mathrm{singlet}}\right|^2$ and $k_{\mathrm{fus}} = k_{\mathrm{fus}} \left|C^{l}_{\mathrm{singlet}}\right|^2$, and of the singlet character $C^{l}_{\mathrm{singlet}}=\left\langle {}^{1}(\hat{T}\hat{T})_0 \middle| \varphi_l \right\rangle$ ($l$ = 1-9) , we computed the eigenstates $\varphi_l$ of the Hamiltonian in Eq. 6 in the main text, and calculated the projections of the nine sublevels on the singlet state ${}^{1}(\hat{T}\hat{T})_0$ in according to $\left|C^{l}_{\mathrm{singlet}}\right|^2$ at different magnetic fields (Figure S1):

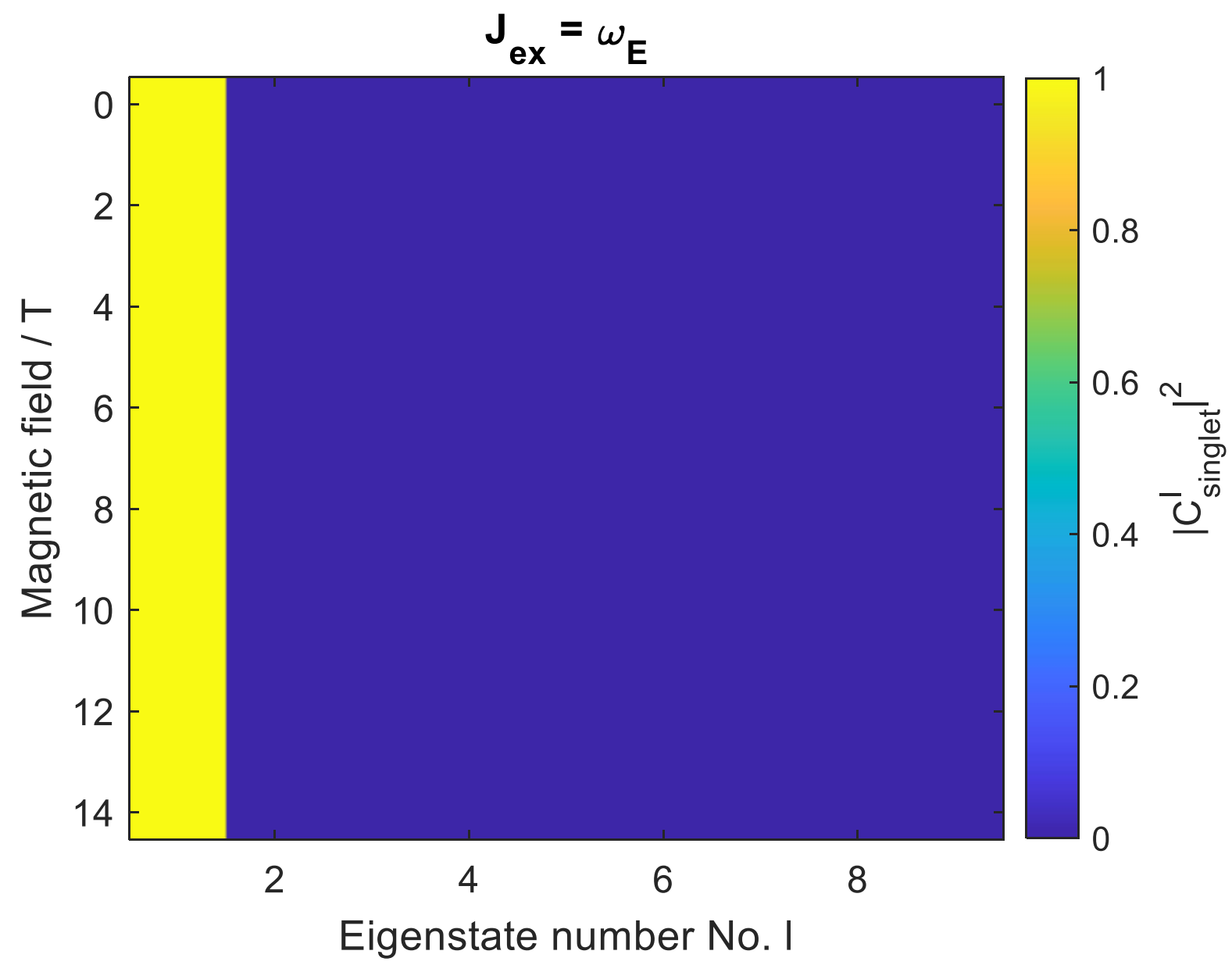

**Figure S1.** Spin projections of the nine triplet pair eigenstates on the singlet state, as a function of the magnetic field.

In the case of strongly coupled triplet pairs ($J_{\mathrm{ex}} \approx$ GHz) and in the absence of the anisotropic part of the spin interactions, the $C^{l}_{\mathrm{singlet}}$ is equal to 1 and the SF kinetics rates are not expected to change with the magnetic field.

## B. Effect of the self-relaxation rates on the DNP enhancement

A difference between the self-relaxation rates of the α and β nuclear components of the triplet pair states, $\Delta R_{self,M/ms}$ is not expected to contribute to the DNP enhancement. The self-relaxation rates of individual triplet pairs have positive and negative magnitude (Figure S2A and S2B), resulting in a difference between the sum of the α and β self-relaxation rates, indicated as $\Delta R_{\mathrm{self,tot}}$, with negative magnitude longer than the timescale of the SF process see Figure S2C:

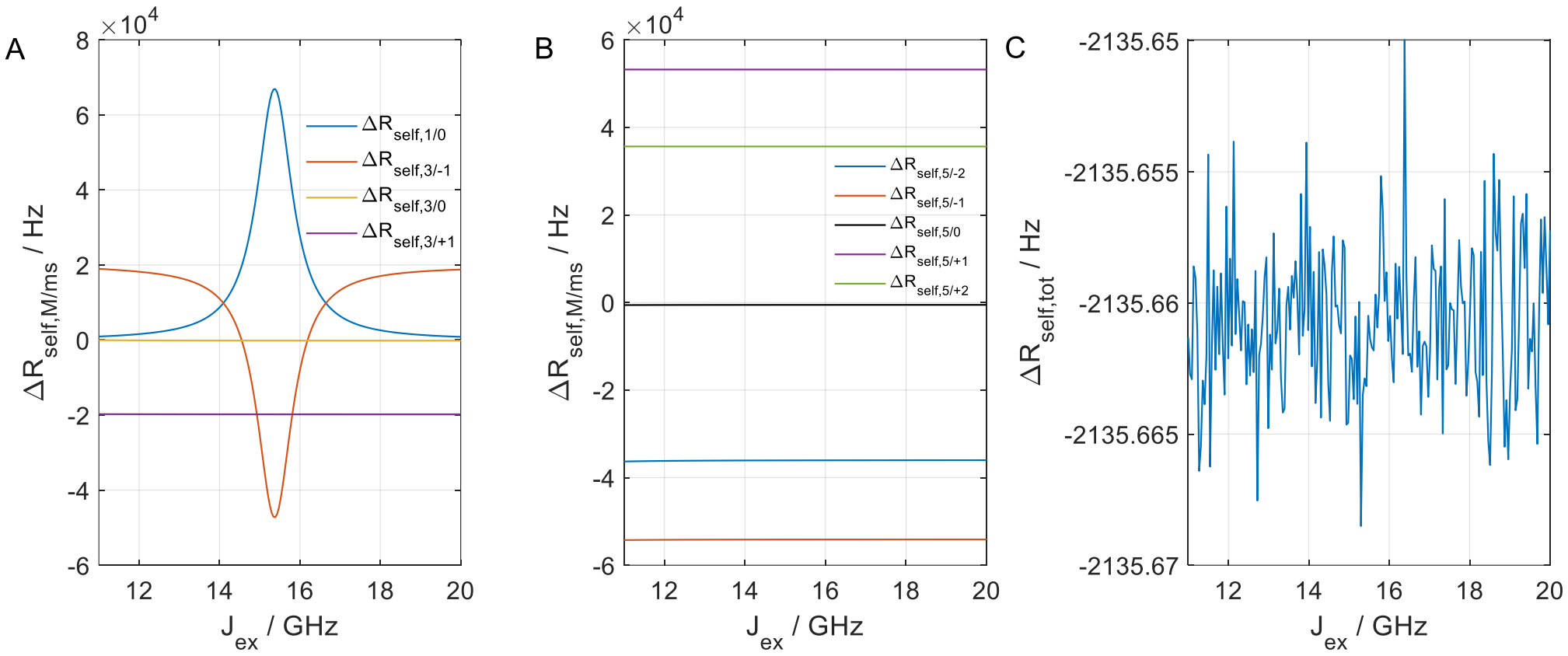

**Figure S2:** (A and B) The difference between the self-relaxation rates of the triplet pair state. (C) The difference between the sum of the self-relaxation rates of the triplet pair states, as a function of the $J_{ex}$. Simulations were performed at 0.55 T. The other simulation parameters are given in the Table 1.

When JDNP condition is fulfilled (at about 15 GHz, see Figure S2), the self-relaxation rates remain constant with the $J_{ex}$ except the states $^1(\hat{T}\hat{T})_0$ and $^3(\hat{T}\hat{T})_{-1}$. The modulations in the self-relaxation rates of $^1(\hat{T}\hat{T})_0$ and $^3(\hat{T}\hat{T})_{-1}$ states are due to the presence of the terms $J(J_{\mathrm{ex}} \mp \omega_{\mathrm{E}} \mp \omega_{\mathrm{N}})$ and $J(J_{\mathrm{ex}} \pm \omega_{\mathrm{E}} \mp \omega_{\mathrm{N}})$ in the spectral density functions, given in Eq. S2 – S10 in the supporting material. $\Delta R_{\mathrm{self,tot}}$ arises mostly from the states $^1(\hat{T}\hat{T})_0$ and $^3(\hat{T}\hat{T})_{-1}$. Eq. S1 shows the spectral density function of $\Delta R_{\mathrm{self,tot}}$ of the triplet pair states:

$$\Delta R_{\mathrm{self,tot}} = \begin{bmatrix} \frac{4}{405}J(J_{\mathrm{ex}} - \omega_{\mathrm{N}}) + \frac{4}{405}J(J_{\mathrm{ex}} + \omega_{\mathrm{N}}) + \frac{1}{81}J(2J_{\mathrm{ex}} - \omega_{\mathrm{N}}) + \frac{1}{81}J(2J_{\mathrm{ex}} + \omega_{\mathrm{N}}) + \\ \frac{8}{405}J(J_{\mathrm{ex}} - \omega_{\mathrm{E}} - \omega_{\mathrm{N}}) + \frac{8}{405}J(J_{\mathrm{ex}} + \omega_{\mathrm{E}} + \omega_{\mathrm{N}}) + \frac{2}{81}J(2J_{\mathrm{ex}} - \omega_{\mathrm{E}} - \omega_{\mathrm{N}}) + \\ \frac{2}{81}J(2J_{\mathrm{ex}} + \omega_{\mathrm{E}} + \omega_{\mathrm{N}}) + \frac{4}{1215}J(J_{\mathrm{ex}} + \omega_{\mathrm{E}} - \omega_{\mathrm{N}}) + \frac{4}{1215}J(J_{\mathrm{ex}} - \omega_{\mathrm{E}} + \omega_{\mathrm{N}}) + \\ + \frac{1}{243}J(2J_{\mathrm{ex}} - \omega_{\mathrm{E}} + \omega_{\mathrm{N}}) + \frac{1}{243}J(2J_{\mathrm{ex}} + \omega_{\mathrm{E}} - \omega_{\mathrm{N}}) \end{bmatrix} \Delta^2_{\mathrm{HFC_1-HFC_2}} + \left[\frac{14}{135}J(\omega_{\mathrm{E}}) + \frac{2}{45}J(\omega_{\mathrm{N}})\right] \Delta^2_{\mathrm{HFC_1+HFC_2}} \tag{S1}$$

where the terms $\Delta^2_{\mathrm{HFC_1+HFC_2}}$ is a matrix corresponding to the second rank norm squared arising from the sum of the anisotropy of the two dipolar hyperfine tensors [1]. $\Delta R_{\mathrm{self,tot}}$ is dominated by the term containing $\Delta^2_{\mathrm{HFC_1+HFC_2}}\, J(\omega_{\mathrm{N}})$ that does not depend on the $J_{ex}$, this explains the stability of $\Delta R_{\mathrm{self,tot}}$ with the $J_{ex}$ in the Figure S2.

### C. Spectral density functions representing the self-relaxation rates:

In the case of ${}^{1}\left(\hat{T}\hat{T}\right)_{0}$, the rate corresponds to:

$$\Delta R_{self,1/0} = R\left[{}^{1}\left(\hat{\mathrm{T}}\hat{\mathrm{T}}\right)_{0,\alpha}\right] - R\left[{}^{1}\left(\hat{\mathrm{T}}\hat{\mathrm{T}}\right)_{0,\beta}\right]$$
$$\approx \frac{32}{135}\left(\aleph_{\mathrm{G1,HFC1}} - \aleph_{\mathrm{G1,HFC2}} - \aleph_{\mathrm{G2,HFC1}} + \aleph_{\mathrm{G2,HFC2}}\right)[J(J_{ex})] +$$
$$\frac{8}{45}\left(\aleph_{\mathrm{G1,HFC1}} - \aleph_{\mathrm{G1,HFC2}} - \aleph_{\mathrm{G2,HFC1}} + \aleph_{\mathrm{G2,HFC2}}\right)[J(J_{ex} - \omega_E) + J(J_{ex} + \omega_E)] +$$
$$\frac{4}{45}\Delta^2_{\mathrm{HFC1-HFC2}}[J(J_{ex} - \omega_E - \omega_N) - J(J_{ex} + \omega_E + \omega_N)] +$$
$$\frac{2}{135}\Delta^2_{\mathrm{HFC1-HFC2}}[J(J_{ex} + \omega_E - \omega_N) - J(J_{ex} - \omega_E + \omega_N)] +$$
$$\frac{2}{45}\Delta^2_{\mathrm{HFC1-HFC2}}[J(J_{ex} - \omega_N) - J(J_{ex} + \omega_N)]$$

(S2)

In the case of ${}^{3}\left(\hat{\mathrm{T}}\hat{\mathrm{T}}\right)_{0}$, the rate corresponds to:

$$\Delta R_{self,3/0} = R\left[{}^{3}\left(\hat{\mathrm{T}}\hat{\mathrm{T}}\right)_{0,\alpha}\right] - R\left[{}^{3}\left(\hat{\mathrm{T}}\hat{\mathrm{T}}\right)_{0,\beta}\right]$$
$$\approx \frac{32}{135}\left(\aleph_{\mathrm{G1,HFC1}} - \aleph_{\mathrm{G1,HFC2}} - \aleph_{\mathrm{G2,HFC1}} + \aleph_{\mathrm{G2,HFC2}}\right)[J(J_{ex})] +$$
$$\frac{16}{135}\left(\aleph_{\mathrm{G1,HFC1}} - \aleph_{\mathrm{G1,HFC2}} - \aleph_{\mathrm{G2,HFC1}} + \aleph_{\mathrm{G2,HFC2}}\right)[J(2J_{ex})] +$$
$$\frac{1}{45}\Delta^2_{\mathrm{HFC1-HFC2}}[J(2J_{ex} - \omega_N) - J(2J_{ex} + \omega_N)] +$$
$$\frac{2}{45}\Delta^2_{\mathrm{HFC1-HFC2}}[J(J_{ex} + \omega_N) - J(J_{ex} - \omega_N)] +$$
$$\frac{1}{15}\begin{pmatrix}\aleph_{\mathrm{G1,HFC1}} - \aleph_{\mathrm{G1,HFC2}} - \aleph_{\mathrm{G2,HFC1}} + \aleph_{\mathrm{G2,HFC2}} + \\ \aleph_{\mathrm{ZFS1,HFC1}} - \aleph_{\mathrm{ZFS2,HFC1}} - \aleph_{\mathrm{ZFS1,HFC2}} + \aleph_{\mathrm{ZFS2,HFC2}}\end{pmatrix}[J(2J_{ex} - \omega_E)] +$$
$$\frac{1}{15}\begin{pmatrix}\aleph_{\mathrm{G1,HFC1}} - \aleph_{\mathrm{G1,HFC2}} - \aleph_{\mathrm{G2,HFC1}} + \aleph_{\mathrm{G2,HFC2}} - \\ \aleph_{\mathrm{ZFS1,HFC1}} + \aleph_{\mathrm{ZFS2,HFC1}} + \aleph_{\mathrm{ZFS1,HFC2}} - \aleph_{\mathrm{ZFS2,HFC2}}\end{pmatrix}[J(2J_{ex} + \omega_E)] +$$
$$\frac{1}{30}\Delta^2_{\mathrm{HFC1-HFC2}}[J(2J_{ex} - \omega_E - \omega_N) - J(2J_{ex} + \omega_E + \omega_N)] +$$
$$\frac{1}{180}\Delta^2_{\mathrm{HFC1-HFC2}}[J(2J_{ex} + \omega_E - \omega_N) - J(2J_{ex} - \omega_E + \omega_N)] + \cdots$$

(S3)

In the case of ${}^{3}\left(\hat{\mathrm{T}}\hat{\mathrm{T}}\right)_{\pm 1}$, the rate corresponds to:

$$\Delta R_{\mathrm{self},3/+1} = R\left[{}^{3}\left(\hat{\mathrm{T}}\hat{\mathrm{T}}\right)_{+1,\alpha}\right] - R\left[{}^{3}\left(\hat{\mathrm{T}}\hat{\mathrm{T}}\right)_{+1,\beta}\right]$$
$$\approx \frac{4}{45}\left(\aleph_{\mathrm{G1,HFC1}} - \aleph_{\mathrm{G1,HFC2}} - \aleph_{\mathrm{G2,HFC1}} + \aleph_{\mathrm{G2,HFC2}}\right)[J(2J_{ex})] +$$
$$\frac{2}{15}\left(\aleph_{\mathrm{ZFS1,HFC1}} - \aleph_{\mathrm{ZFS1,HFC2}} - \aleph_{\mathrm{ZFS2,HFC1}} + \aleph_{\mathrm{ZFS2,HFC2}}\right)[J(2J_{ex})] +$$
$$\frac{1}{45}\begin{pmatrix}\aleph_{\mathrm{G1,HFC1}} - \aleph_{\mathrm{G1,HFC2}} - \aleph_{\mathrm{G2,HFC1}} + \aleph_{\mathrm{G2,HFC2}} + \\ 3\aleph_{\mathrm{ZFS1,HFC1}} - 3\aleph_{\mathrm{ZFS1,HFC2}} - 3\aleph_{\mathrm{ZFS2,HFC1}} + 3\aleph_{\mathrm{ZFS2,HFC2}}\end{pmatrix}[J(2J_{ex} - \omega_E)] +$$
$$\frac{2}{15}\begin{pmatrix}\aleph_{\mathrm{G1,HFC1}} - \aleph_{\mathrm{G1,HFC2}} - \aleph_{\mathrm{G2,HFC1}} + \aleph_{\mathrm{G2,HFC2}} + \\ \aleph_{\mathrm{ZFS1,HFC1}} - \aleph_{\mathrm{ZFS1,HFC2}} - \aleph_{\mathrm{ZFS2,HFC1}} + \aleph_{\mathrm{ZFS2,HFC2}}\end{pmatrix}[J(2J_{ex} + \omega_E)] +$$

$$\frac{8}{45}\left(\aleph_{\mathrm{G1,HFC1}}-\aleph_{\mathrm{G1,HFC2}}-\aleph_{\mathrm{G2,HFC1}}+\aleph_{\mathrm{G2,HFC2}}\right)[J(J_{ex}+\omega_E)]+$$
$$\frac{1}{90}\Delta^2_{\mathrm{HFC1-HFC2}}[J(2J_{ex}-\omega_E-\omega_N)] - \frac{1}{15}\Delta^2_{\mathrm{HFC1-HFC2}}[J(2J_{ex}+\omega_E+\omega_N)]+$$
$$\frac{1}{90}\Delta^2_{\mathrm{HFC1-HFC2}}[J(2J_{ex}+\omega_E-\omega_N)] - \frac{1}{540}\Delta^2_{\mathrm{HFC1-HFC2}}[J(2J_{ex}-\omega_E+\omega_N)]+$$
$$\frac{4}{45}\Delta^2_{\mathrm{HFC1-HFC2}}[J(J_{ex}+\omega_E+\omega_N)] - \frac{2}{135}\Delta^2_{\mathrm{HFC1-HFC2}}[J(J_{ex}+\omega_E-\omega_N)]+$$
$$\frac{1}{60}\Delta^2_{\mathrm{HFC1-HFC2}}[J(2J_{ex}-\omega_N)-J(2J_{ex}+\omega_N)]+\ldots$$

(S4)

and:

$$\Delta R_{\mathrm{self},3/-1} = R\left[{}^3(\widehat{\mathrm{T}}\widehat{\mathrm{T}})_{-1,\alpha}\right] - R\left[{}^3(\widehat{\mathrm{T}}\widehat{\mathrm{T}})_{-1,\beta}\right]$$

$$\approx \frac{4}{45}\left(\aleph_{\mathrm{G1,HFC1}}-\aleph_{\mathrm{G1,HFC2}}-\aleph_{\mathrm{G2,HFC1}}+\aleph_{\mathrm{G2,HFC2}}\right)[J(2J_{ex})]+$$
$$\frac{2}{15}\left(-\aleph_{\mathrm{ZFS1,HFC1}}+\aleph_{\mathrm{ZFS1,HFC2}}+\aleph_{\mathrm{ZFS2,HFC1}}-\aleph_{\mathrm{ZFS2,HFC2}}\right)[J(2J_{ex})]+$$
$$\frac{1}{45}\begin{pmatrix}\aleph_{\mathrm{G1,HFC1}}-\aleph_{\mathrm{G1,HFC2}}-\aleph_{\mathrm{G2,HFC1}}+\aleph_{\mathrm{G2,HFC2}}- \\ 3\aleph_{\mathrm{ZFS1,HFC1}}+3\aleph_{\mathrm{ZFS1,HFC2}}+3\aleph_{\mathrm{ZFS2,HFC1}}-3\aleph_{\mathrm{ZFS2,HFC2}}\end{pmatrix}[J(2J_{ex}+\omega_E)]+$$
$$\frac{2}{15}\begin{pmatrix}\aleph_{\mathrm{G1,HFC1}}-\aleph_{\mathrm{G1,HFC2}}-\aleph_{\mathrm{G2,HFC1}}+\aleph_{\mathrm{G2,HFC2}}- \\ \aleph_{\mathrm{ZFS1,HFC1}}+\aleph_{\mathrm{ZFS1,HFC2}}+\aleph_{\mathrm{ZFS2,HFC1}}-\aleph_{\mathrm{ZFS2,HFC2}}\end{pmatrix}[J(2J_{ex}-\omega_E)]+$$
$$\frac{8}{45}\left(\aleph_{\mathrm{G1,HFC1}}-\aleph_{\mathrm{G1,HFC2}}-\aleph_{\mathrm{G2,HFC1}}+\aleph_{\mathrm{G2,HFC2}}\right)[J(J_{ex}-\omega_E)]-$$
$$\frac{1}{90}\Delta^2_{\mathrm{HFC1-HFC2}}[J(2J_{ex}-\omega_E+\omega_N)] + \frac{1}{15}\Delta^2_{\mathrm{HFC1-HFC2}}[J(2J_{ex}-\omega_E-\omega_N)]-$$
$$\frac{1}{90}\Delta^2_{\mathrm{HFC1-HFC2}}[J(2J_{ex}+\omega_E+\omega_N)] - \frac{1}{540}\Delta^2_{\mathrm{HFC1-HFC2}}[J(2J_{ex}+\omega_E-\omega_N)]-$$
$$\frac{4}{45}\Delta^2_{\mathrm{HFC1-HFC2}}[J(J_{ex}-\omega_E-\omega_N)] + \frac{2}{135}\Delta^2_{\mathrm{HFC1-HFC2}}[J(J_{ex}-\omega_E+\omega_N)]+$$
$$\frac{1}{60}\Delta^2_{\mathrm{HFC1-HFC2}}[J(2J_{ex}-\omega_N)-J(2J_{ex}+\omega_N)]+\ldots$$

(S5)

In the case of ${}^5(\widehat{\mathrm{T}}\widehat{\mathrm{T}})_{\pm 1}$, the rate corresponds to:

$$\Delta R_{\mathrm{self},5/+1} = R\left[{}^5(\widehat{\mathrm{T}}\widehat{\mathrm{T}})_{+1,\alpha}\right] - R\left[{}^5(\widehat{\mathrm{T}}\widehat{\mathrm{T}})_{+1,\beta}\right]$$

$$\approx \frac{4}{45}\left(\aleph_{\mathrm{G1,HFC1}}-\aleph_{\mathrm{G1,HFC2}}-\aleph_{\mathrm{G2,HFC1}}+\aleph_{\mathrm{G2,HFC2}}\right)[J(2J_{ex})]+$$
$$\frac{2}{15}\left(\aleph_{\mathrm{ZFS1,HFC1}}-\aleph_{\mathrm{ZFS1,HFC2}}-\aleph_{\mathrm{ZFS2,HFC1}}+\aleph_{\mathrm{ZFS2,HFC2}}\right)[J(2J_{ex})]+$$
$$\frac{1}{60}\Delta^2_{\mathrm{HFC1-HFC2}}[J(2J_{ex}+\omega_N)-J(2J_{ex}-\omega_N)] +$$
$$\frac{1}{30}\Delta^2_{\mathrm{HFC1-HFC2}}[J(2J_{ex}+\omega_E+\omega_N)] - \frac{1}{180}\Delta^2_{\mathrm{HFC1-HFC2}}[J(2J_{ex}+\omega_E-\omega_N)]+$$
$$\frac{1}{15}\begin{pmatrix}\aleph_{\mathrm{G1,HFC1}}-\aleph_{\mathrm{G1,HFC2}}-\aleph_{\mathrm{G2,HFC1}}+\aleph_{\mathrm{G2,HFC2}}- \\ \aleph_{\mathrm{ZFS1,HFC1}}+\aleph_{\mathrm{ZFS1,HFC2}}+\aleph_{\mathrm{ZFS2,HFC1}}-\aleph_{\mathrm{ZFS2,HFC2}}\end{pmatrix}[J(2J_{ex}+\omega_E)]+\ldots$$

(S6)

and:

$$\Delta R_{\mathrm{self},5/-1} = R\left[{}^5(\widehat{\mathrm{T}}\widehat{\mathrm{T}})_{-1,\alpha}\right] - R\left[{}^5(\widehat{\mathrm{T}}\widehat{\mathrm{T}})_{-1,\beta}\right]$$

$$\approx \frac{4}{45}\left(\aleph_{\mathrm{G1,HFC1}}-\aleph_{\mathrm{G1,HFC2}}-\aleph_{\mathrm{G2,HFC1}}+\aleph_{\mathrm{G2,HFC2}}\right)[J(2J_{ex})]+$$

$$\frac{2}{15}\left(-\aleph_{\mathrm{ZFS1,HFC1}}+\aleph_{\mathrm{ZFS1,HFC2}}+\aleph_{\mathrm{ZFS2,HFC1}}-\aleph_{\mathrm{ZFS2,HFC2}}\right)[J(2J_{ex})]+$$
$$\frac{1}{60}\Delta^2_{\mathrm{HFC1-HFC2}}[J(2J_{ex}+\omega_N)-J(2J_{ex}-\omega_N)]\ -$$
$$\frac{1}{30}\Delta^2_{\mathrm{HFC1-HFC2}}[J(2J_{ex}-\omega_E-\omega_N)]+\frac{1}{180}\Delta^2_{\mathrm{HFC1-HFC2}}[J(2J_{ex}-\omega_E+\omega_N)]+$$
$$\frac{1}{15}\begin{pmatrix}\aleph_{\mathrm{G1,HFC1}}-\aleph_{\mathrm{G1,HFC2}}-\aleph_{\mathrm{G2,HFC1}}+\aleph_{\mathrm{G2,HFC2}}+\\ \aleph_{\mathrm{ZFS1,HFC1}}-\aleph_{\mathrm{ZFS1,HFC2}}-\aleph_{\mathrm{ZFS2,HFC1}}+\aleph_{\mathrm{ZFS2,HFC2}}\end{pmatrix}[J(2J_{ex}-\omega_E)]+\ldots$$

(S7)

In the case of ${}^5\left(\widehat{\mathrm{T}}\widehat{\mathrm{T}}\right)_0$, the rate corresponds to:

$$\Delta R_{\mathrm{self},5/0}=R\left[{}^5\left(\widehat{\mathrm{T}}\widehat{\mathrm{T}}\right)_{0,\alpha}\right]-\ R\left[{}^5\left(\widehat{\mathrm{T}}\widehat{\mathrm{T}}\right)_{0,\beta}\right]$$
$$\approx\frac{16}{135}\left(\aleph_{\mathrm{G1,HFC1}}-\aleph_{\mathrm{G1,HFC2}}-\aleph_{\mathrm{G2,HFC1}}+\aleph_{\mathrm{G2,HFC2}}\right)[J(2J_{ex})]+$$
$$\frac{1}{45}\Delta^2_{\mathrm{HFC1-HFC2}}[J(2J_{ex}+\omega_N)-J(2J_{ex}-\omega_N)]\ +$$
$$\frac{1}{540}\Delta^2_{\mathrm{HFC1-HFC2}}[J(2J_{ex}-\omega_E+\omega_N)-J(2J_{ex}+\omega_E-\omega_N)]\ +$$
$$\frac{1}{90}\Delta^2_{\mathrm{HFC1-HFC2}}[J(2J_{ex}+\omega_E+\omega_N)-J(2J_{ex}-\omega_E-\omega_N)]+$$
$$\frac{1}{45}\begin{pmatrix}\aleph_{\mathrm{G1,HFC1}}-\aleph_{\mathrm{G1,HFC2}}-\aleph_{\mathrm{G2,HFC1}}+\aleph_{\mathrm{G2,HFC2}}-\\ 3\aleph_{\mathrm{ZFS1,HFC1}}+3\aleph_{\mathrm{ZFS1,HFC2}}+3\aleph_{\mathrm{ZFS2,HFC1}}-3\aleph_{\mathrm{ZFS2,HFC2}}\end{pmatrix}[J(2J_{ex}+\omega_E)]+$$
$$\frac{1}{45}\begin{pmatrix}\aleph_{\mathrm{G1,HFC1}}-\aleph_{\mathrm{G1,HFC2}}-\aleph_{\mathrm{G2,HFC1}}+\aleph_{\mathrm{G2,HFC2}}+\\ 3\aleph_{\mathrm{ZFS1,HFC1}}-3\aleph_{\mathrm{ZFS1,HFC2}}-3\aleph_{\mathrm{ZFS2,HFC1}}+3\aleph_{\mathrm{ZFS2,HFC2}}\end{pmatrix}[J(2J_{ex}-\omega_E)]+\ldots$$

(S8)

In the case of ${}^5\left(\hat{T}\hat{T}\right)_{\pm 2}$, the rate corresponds to:

$$\Delta R_{\mathrm{self},5/+2}=R\left[{}^5\left(\widehat{\mathrm{T}}\widehat{\mathrm{T}}\right)_{+2,\alpha}\right]-\ R\left[{}^5\left(\widehat{\mathrm{T}}\widehat{\mathrm{T}}\right)_{+2,\beta}\right]$$
$$\approx\frac{1}{15}\Delta^2_{\mathrm{HFC1-HFC2}}[J(2J_{\mathrm{ex}}+\omega_{\mathrm{E}}+\omega_{\mathrm{N}})]-\frac{1}{90}\Delta^2_{\mathrm{HFC1-HFC2}}[J(2J_{\mathrm{ex}}+\omega_{\mathrm{E}}-\omega_{\mathrm{N}})]+$$
$$\frac{2}{15}\begin{pmatrix}\aleph_{\mathrm{G1,HFC1}}-\aleph_{\mathrm{G1,HFC2}}-\aleph_{\mathrm{G2,HFC1}}+\aleph_{\mathrm{G2,HFC2}}+\\ \aleph_{\mathrm{ZFS1,HFC1}}-\aleph_{\mathrm{ZFS1,HFC2}}-\aleph_{\mathrm{ZFS2,HFC1}}+\aleph_{\mathrm{ZFS2,HFC2}}\end{pmatrix}[J(2J_{\mathrm{ex}}+\omega_{\mathrm{E}})]+\cdots$$

(S9)

and:

$$\Delta R_{\mathrm{self},5/-2}=R\left[{}^5\left(\widehat{\mathrm{T}}\widehat{\mathrm{T}}\right)_{-2,\alpha}\right]-\ R\left[{}^5\left(\widehat{\mathrm{T}}\widehat{\mathrm{T}}\right)_{-2,\beta}\right]$$
$$\approx-\frac{1}{15}\Delta^2_{\mathrm{HFC1-HFC2}}[J(2J_{ex}-\omega_E-\omega_N)]+\ \frac{1}{90}\Delta^2_{\mathrm{HFC1-HFC2}}[J(2J_{ex}-\omega_E+\omega_N)]+$$
$$\frac{2}{15}\begin{pmatrix}\aleph_{\mathrm{G1,HFC1}}-\aleph_{\mathrm{G1,HFC2}}-\aleph_{\mathrm{G2,HFC1}}+\aleph_{\mathrm{G2,HFC2}}-\\ \aleph_{\mathrm{ZFS1,HFC1}}+\aleph_{\mathrm{ZFS1,HFC2}}+\aleph_{\mathrm{ZFS2,HFC1}}-\aleph_{\mathrm{ZFS2,HFC2}}\end{pmatrix}[J(2J_{ex}-\omega_E)]+\ldots$$

(S10)

where the $\Delta^2_{\mathbf{A}}$ is a matric corresponding to the second rank norm squared [1] that arises from the anisotropy of a tensor **A**; $\aleph_{\mathbf{A,B}}$ is a matrix corresponding to the second-rank scalar products of the tensors **A** and **B**. It arises from anisotropies associated to the electron Zeeman interaction tensor G, the zero-filed splitting tensor, ZFS and the inter-electron dipolar interaction tensor, DD of the triplet 1 and 2 respectively. The terms containing $J(\omega_E)$ do not contribute to the DNP enhancement and are omitted for simplicity.

### D. Effects of the kinetic rates and of the $J_{ex}$ on the DNP enhancement

The effect of the kinetics rates, as a function of the $J_{ex}$ on a hypothetical pentacene dimer molecule that can fulfill the JDNP condition ($J_{ex} \approx \omega_E$ = 263 GHz) at a high magnet field (e.g. 9.4 T) is shown in Figure S3. The strongest DNP enhancements were observed at the condition $J_{ex} \approx \omega_E$ = 263 GHz ± 1 GHz.

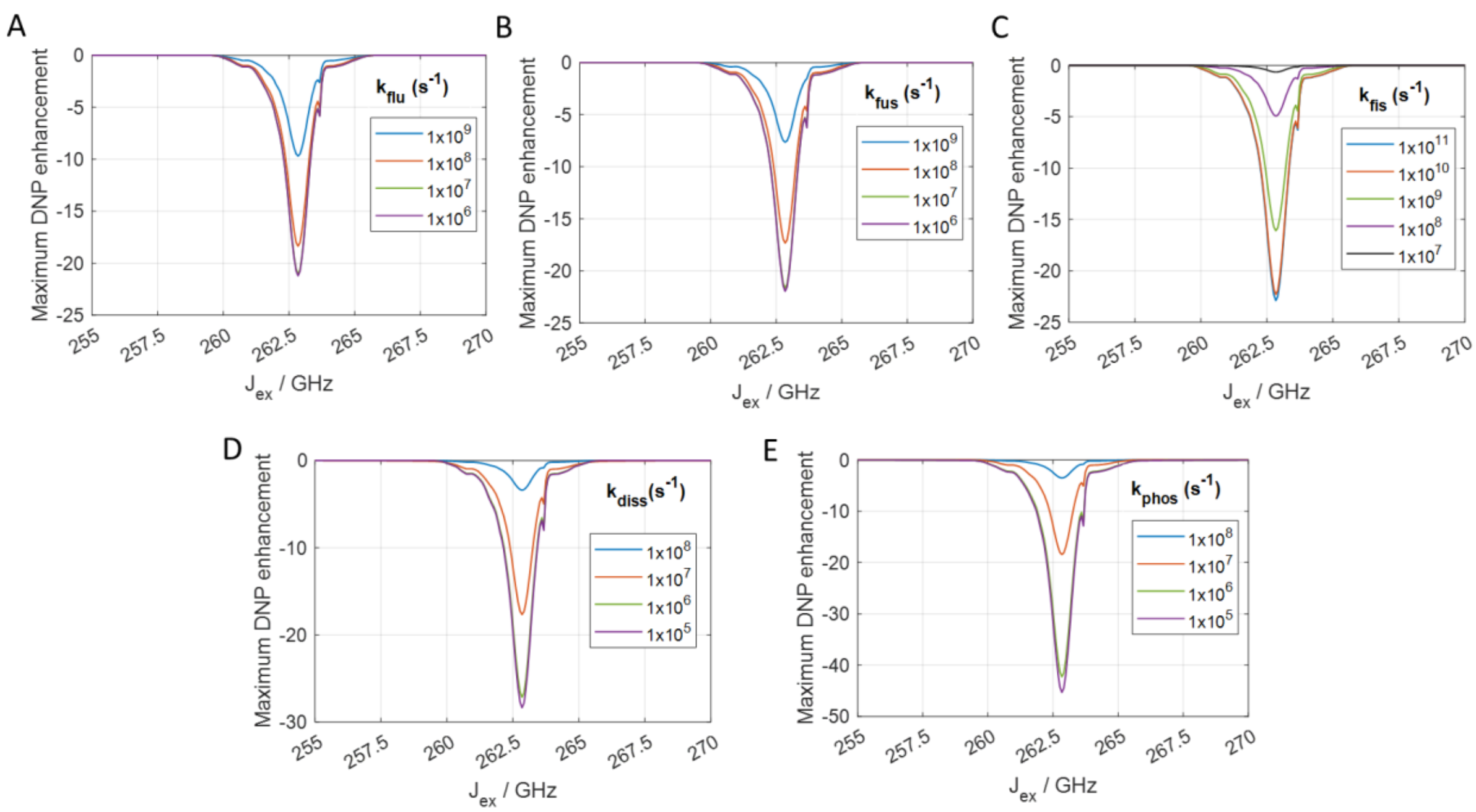

**Figure S3.** Effects of the kinetic rates and of the $J_{ex}$ on the DNP enhancement: (A) $k_{flu}$, (B) $k_{fus}$, (C) $k_{fis}$, (D) $k_{diss}$ and (E) $k_{phos}$. Simulations were performed at 9.4 T. While the kinetics rate is changed, the other rates and parameters were kept the same as those shown in Table 1.

**E. Effect of anisotropic *g*-tensors on the self-relaxation rates and on the DNP enhancement:**

As discussed previously, [2, 3] only rotations about the linker connecting the two monomeric units in a symmetric dimer (also in the case of triplet-pairs possessing nearly isotropic or axially symmetric *g*-tensors) can lead to a difference between the two anisotropic *g*- and ZFS-tensors. As discussed in the supporting information of [3], the β bending angle (about the linker connecting the two monomeric units) in long (~38 Å) linear linkers containing phenylene and acetylene units ranges between about 0° and ±40°. Smaller displacements were expected for shorter linkers (~14 Å – 20 Å) as in the pentacene dimers considered for triplet-JDNP. Therefore, we calculated the DNP enhancements considering β and β' angles set equal to 0° and 22°, at 0.55 T, 9.4 T and 14.1 T, see Figure S4.

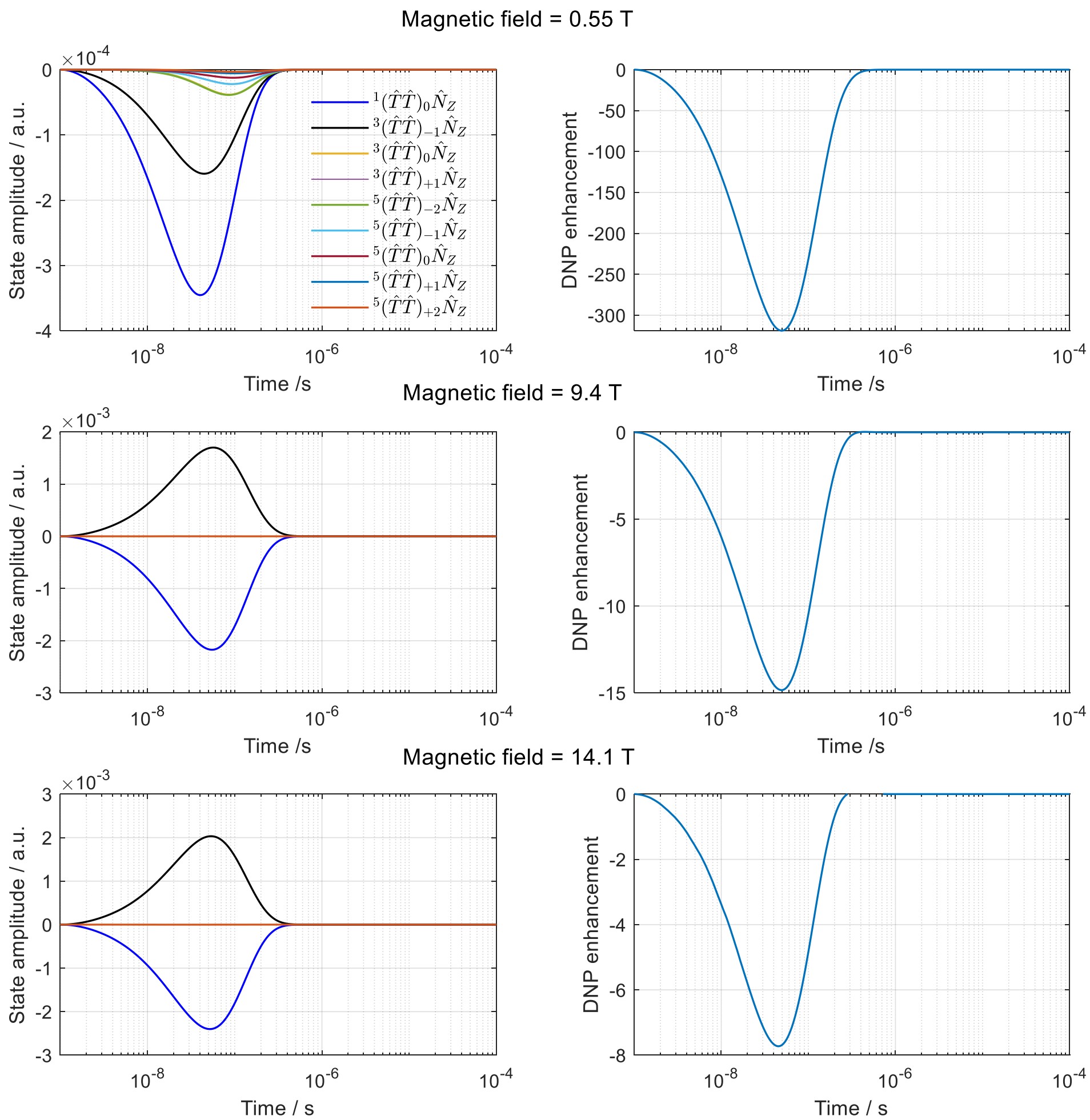

**Figure S4**. Time domain simulations performed at 0.55 T, 9.4 T and 14.1 T, showing the imbalance between the α and β components of the triplet pair states (on the left) and the resulting DNP enhancement (on the right). The simulation parameters were set the same as those used in Figure 2 in the main text, but the β and β' angles were set to 0° and +22°, while the angles α (α') and γ (γ') were set to zero.

In Figure S4, we observed a decrease of the DNP enhancement due to the decreased amplitude of $^3(\hat{T}\hat{T})_{-1}\hat{N}_Z$ that changes sign at magnetic fields above 0 T. The magnitude of the self- and cross- relaxation rates of the states $^3(\hat{T}\hat{T})_0$ and $^3(\hat{T}\hat{T})_{-1}$ involved in the polarization process do not change in the case of non-collinear $g$-tensors. However, we observed a change in the sign of the rate $\Delta R_{\mathrm{self},3/-1}$ that led to a change in $\Delta R_{\mathrm{self,tot}}$ (see Figure S5).

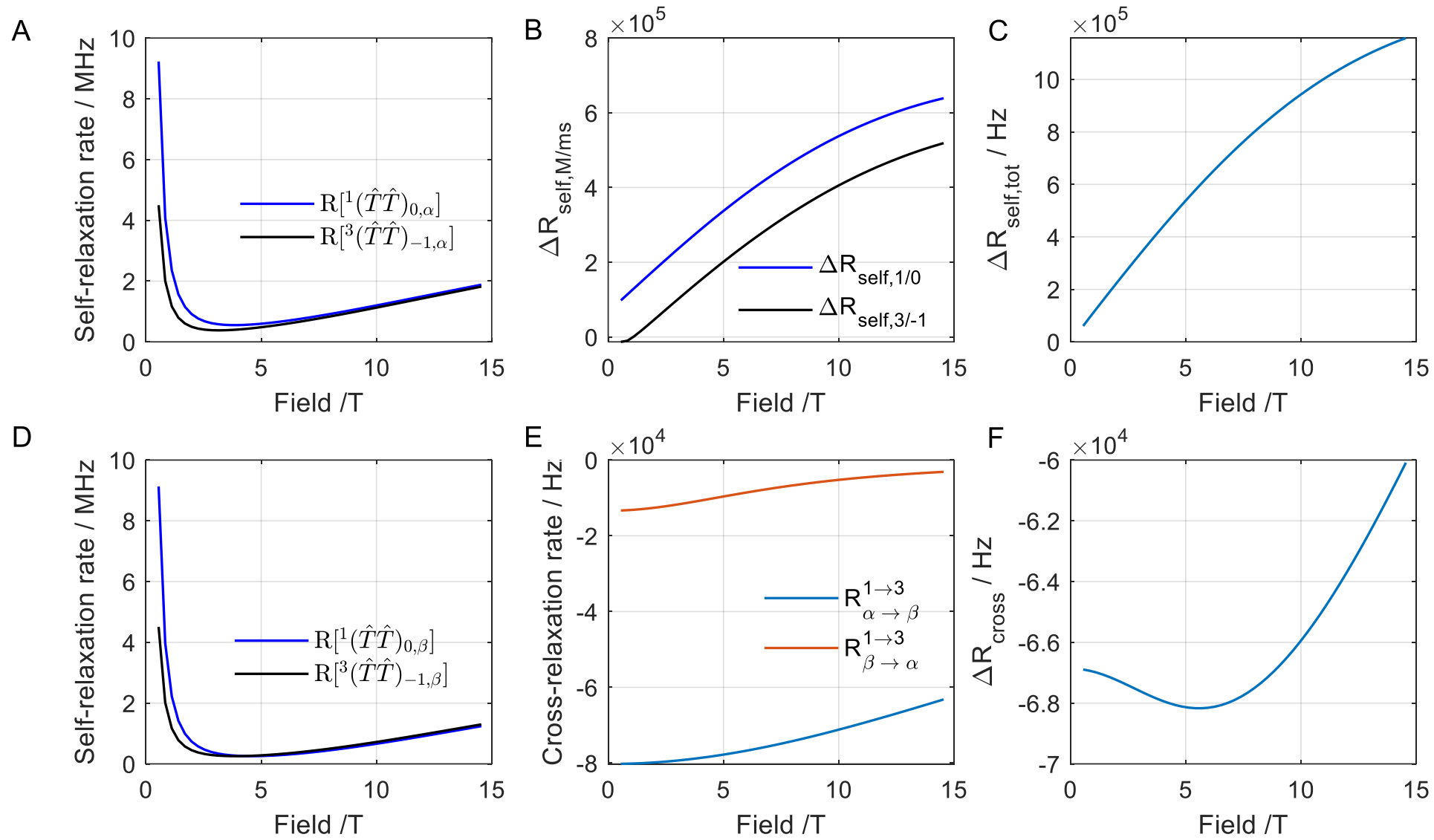

**Figure S5:** (A and D) Self-relaxation rates of the α and β nuclear components of the triplet pair states $^3(\hat{T}\hat{T})_0$ and $^3(\hat{T}\hat{T})_{-1}$, (B) $\Delta R_{\mathrm{self,M/ms}}$ of the triplet pair states $^3(\hat{T}\hat{T})_0$ and $^3(\hat{T}\hat{T})_{-1}$, (C) the difference in self-relaxation rates $\Delta R_{\mathrm{self,tot}}$, (E) cross-relaxation rates $R^{1\to 3}_{\alpha\to\beta}$ and $R^{1\to 3}_{\beta\to\alpha}$, (F) the difference in cross-relaxation rates $\Delta R_{\mathrm{cross}} = R^{1\to 3}_{\alpha\to\beta} - R^{1\to 3}_{\beta\to\alpha}$, as a function of the magnetic field. The simulation parameters were set the same as those used in Figure 4 in the main text, but the β and β' angles were set to 0° and +22°, while the angles α (α') and γ (γ') were set to zero.

The change of the sign of $\Delta R_{\mathrm{self},3/-1}$ led to a change in the sign and amplitude of $^3(\hat{T}\hat{T})_{-1}\hat{N}_Z$ at magnetic fields above 0 T. It does not result in a change of the sign of the overall DNP enhancement, but just in a small reduction of it, see Figure S6.

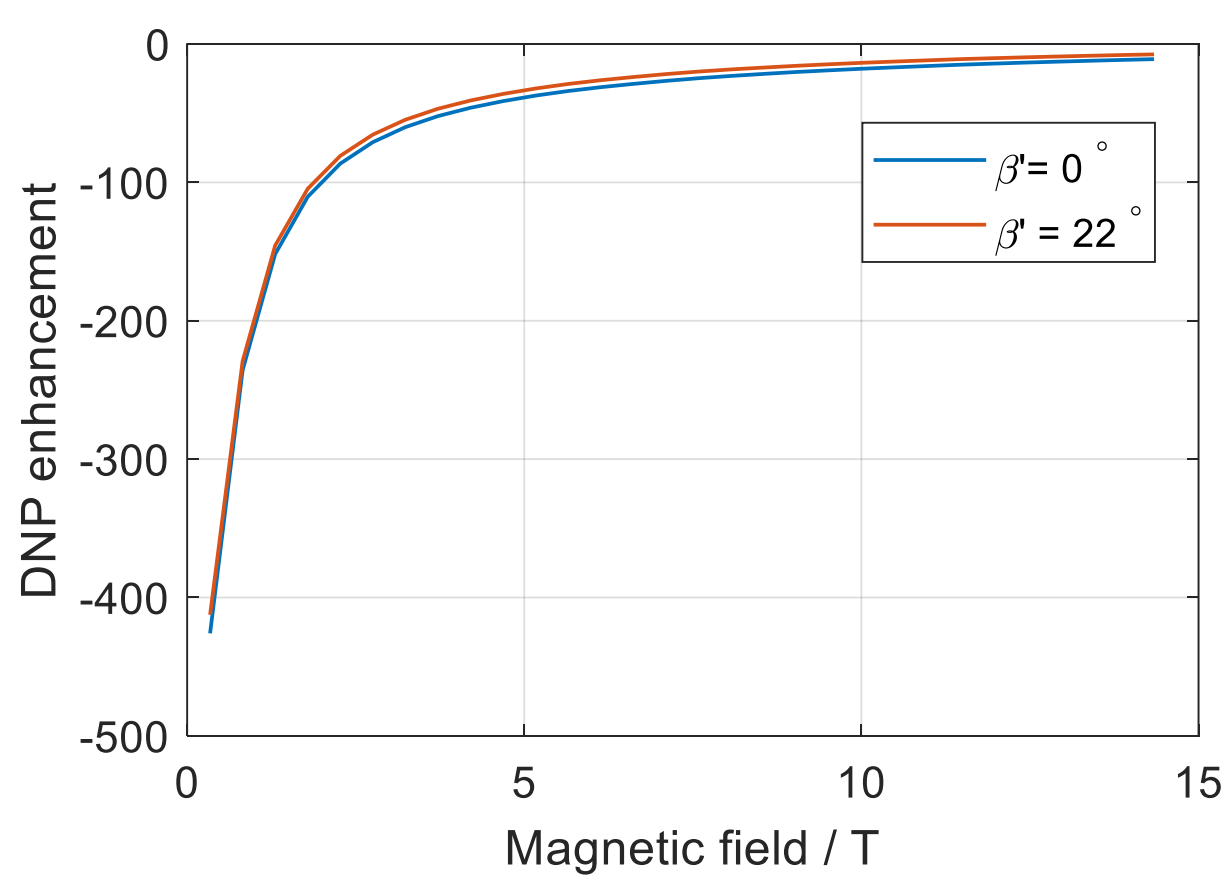

**Figure S6.** Comparison between the maximum enhancement obtained in the triplet-JDNP time domain simulations as a function of the magnetic field using the simulation parameters are given in Table 1 with the β and β' angles were set to 0° and 0° (blue line) and to 0° and +22° (red line), while the angles α (α') and γ (γ') were set to zero. The $J_{ex}$ was set at the condition $J_{ex} \approx \omega_E$.

### F. Effects of the inter-electron distance and the rotational correlation time on the DNP enhancement

DNP enhancement does not change with the variation of the inter-electron distance in TIPS-pentacene dimers bridged by phenylenes / acetylene linkers, like those shown in Figure 8 in the main text, see Figure 7A. The variation of the inter-electron distance would lead only to a slight variation in the electron relaxation rates. We observed an increase of the DNP enhancement with the rotational correlation times of dipolar hyperfine coupled triad and with the magnetic field, see Figure S7B. This arises from an increase of the rates $R^{1\rightarrow3}_{\alpha\rightarrow\beta}$ and $R^{1\rightarrow3}_{\beta\rightarrow\alpha}$, and of their difference with the rotational correlation time. DNP enhancements above ≥5 can be obtained at high magnetic fields between 9 T and 14 T with rotational correlation times ≥150 ps.

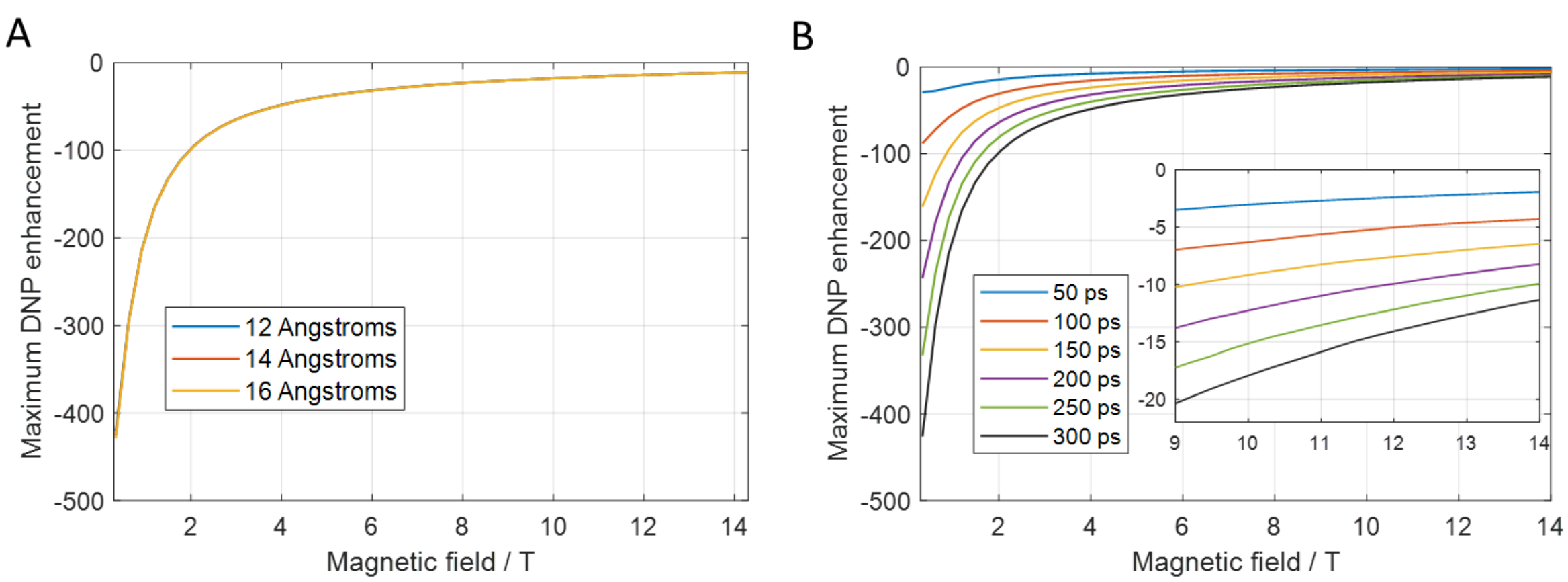

**Figure S7:** Maximum enhancement obtained in the triplet-JDNP time-domain simulations as a function of in the inter-electron distance (A) and the rotational correlation time (B), and the magnetic field. The $J_{ex}$ was set at the JDNP conditions $J_{ex} \approx \omega_E$, and the other simulation parameters are given in Table 1.